# Morphological, nanostructural, and compositional evolution during phase separation of a model Ni-Al-Mo superalloy: Atom-probe tomographic experiments and lattice-kinetic Monte Carlo simulations

Yiyou Tu[1,2], Zugang Mao[2], Ronald D. Noebe,[3]   David N. Seidman[2,4,*]

[1]School of Materials Science and Engineering, Southeast University, Jiyin Road, Jiangning District, Nanjing, Jiangsu 211189, China
[2]Department of Materials Science and Engineering, Northwestern University, 2220 Campus Drive, Evanston, IL 60208-3108, USA
[3]NASA Glenn Research Center, Materials and Structures Division, 21000 Brookpark Road, Cleveland, OH 44135-3191, USA
[4]Northwestern University Center for Atom-Probe Tomography (NUCAPT), 2220 Campus Drive, Evanston, IL 60208-3108, USA

## Abstract

The details of phase separation of a Ni-6.5Al-9.9Mo at. % aged at 978 K (705 ºC) for aging times ranging from 1/6 to 1024 h are investigated by atom-probe tomography and lattice-kinetic Monte Carlo (LKMC) simulations. On the basis of the temporal evolution of the nanostructure, three experimental regimes are identified: (1) concomitant precipitate nucleation and growth (t <1/4 h); (2) concurrent coagulation and coalescence (t = 1/4 to 16 h); and (3) quasi-stationary coarsening of γ'(L1$_2$)-precipitates (t = 16 to 1024 h). The temporal dependencies of the mean precipitate radius, $<R(t)>$, and precipitate number density, $N_v(t)$, are determined experimentally,

[*]Corresponding author at: Department of Materials Science and Engineering, Northwestern University, 2220 Campus Drive, Evanston, IL 60208-3108, USA.

Tel.: +1 847 491 4391; fax: +1 847 491 7820.

E-mail address: d-seidman@northwestern.edu (D.N. Seidman).

0.344 ± 0.012 and -0.95±0.02, respectively, following the predictions of quasi-stationary coarsening models. In this alloy aged at 978 K (705 °5), Al partitions strongly to the γ'(L1$_2$)-phase with a partitioning coefficient $\kappa_{Al}$ = 4.06±0.04, whereas Mo and Ni partition to the γ(f.c.c.)-matrix with $\kappa_i$ values of 0.61±0.01 and 0.90±0.01, respectively. In the quasi-stationary regime (t > 16 h), the temporal exponents of the Al, Mo, and Ni supersaturations in both the γ(f.c.c.)-matrix and γ'(L1$_2$)-precipitates are in reasonable agreement with a multi-component coarsening model's prediction of -1/3. Quantitative analyses of the edge-to-edge inter-precipitate distances demonstrate that coagulation and coalescence are consequences of the overlap of the diffusion fields surrounding the γ'(L1$_2$)-precipitates. Both 3-D APT and LKMC results demonstrate that the interfacial compositional width, $\delta_i(t)$, and $\delta_i(t)/<R(t)>$ values decrease with increasing <R(t)> values. And the interfacial compositional width at infinite ageing time, $\delta_i(t)$, are estimated to be 1.89±0.22 nm, 2.09±0.12 nm and 2.64±0.03 nm for Ni, Al and Mo, respectively.

## 1. Introduction

Nickel-based superalloys have been studied intensively for many years because of their excellent high-temperature strength and their resistance against coarsening, creep and corrosion [1, 2]. Their high-temperature strength and creep resistance are primarily the result of an ordered phase of hard coherent γ′(L1$_2$) precipitates embedded in a disordered γ(f.c.c.)-matrix, which behave as hard obstacles to dislocation motion; the alloys can accommodate substantial concentrations of refractory elemental additions, such as Cr, Mo, W, Ta, Re, Ru, and Hf   [2-5]. The kinetic pathways that lead to phase separation at service temperatures must be understood quantitatively to improve their properties through process optimization and to develop reliable life-time prediction models [6].

The precipitation (or phase separation) of the γ′(L1$_2$)-phase from a supersaturated γ(f.c.c.)-matrix of nickel-based superalloys has been investigated by conventional and high-resolution transmission electron microscopy (TEM) [7-9], X-ray analysis[10], small-angle and wide-angle neutron scattering [11-14], atom-probe field-ion microscopy (APFIM) [15, 16], atom-probe tomography (APT) [17-24], phase-field modeling [25-27], and lattice kinetic Monte Carlo (LKMC) simulations [28, 29]. In view of the subnanometer scale of spatial and analytical resolutions for compositions, APT is the most effective method for understanding phase separation of multi-component superalloys [30]. Combining APT and LKMC simulations is a powerful approach for understanding phase separation on an atomic scale, particularly the early-stage transient occurrence of coagulating γ′(L1$_2$)-precipitates [28, 29]. The kinetic correlations among different atomic fluxes in the simulations are *modified by changing the vacancy-solute interactions* out to fourth nearest-neighbor (NN) distances, resulting in a suppression of coagulation, which demonstrates that the diffusion mechanism strongly affects the kinetic pathways for phase-separation. Despite having many qualitatively correct predictions, the mean-field approach for the phase separation of a ternary alloy can fail to describe quantitatively the kinetic pathways that lead to phase separation in concentrated metallic alloys if it neglects flux couplings [29].

Similar to Cr [31, 32], Mo partitions to and increases the lattice parameter of the $\gamma$(f.c.c.)-matrix in two-phase Ni-based alloys. Therefore, controlled ternary additions of Mo to binary Ni-Al alloys results in variations of the lattice parameter misfit between the $\gamma$(f.c.c.)-solid-solution and the $\gamma'$(L1$_2$)-ordered-precipitates. The addition of Mo to the binary Ni–Al system reduces the lattice parameter misfit between the $\gamma'$(L1$_2$)-Ni$_3$(Al$_x$Mo$_{1-x}$)-precipitates and the $\gamma$(f.c.c.)-matrix, often leading to $\gamma'$(L1$_2$)-precipitates, which are nearly misfit-free and ultimately allowing the $\gamma'$(L1$_2$)-precipitates to remain spheroidal with large dimensions, that is,~100 nm with increasing aging time; in the present case out to 1024 h [33, 34].

Ni-Al-Mo alloys are excellent candidates for combined APT experimental data and LKMC simulations with the predictions of nucleation, growth, and coarsening for multi-component alloys because of the coherent, spheroidal $\gamma'$(L1$_2$)-precipitates with relatively stress-free matrix/precipitate heterophase interfaces. Efforts to understand the phase separation of a Ni-Al-Mo ternary alloy have created a demand for a quantitative investigation of the $\gamma'$(L1$_2$)-precipitate-phase's morphological development, phase composition, and nanostructural evolution [35, 36]. We focus on characterizing the morphological, nanostructural evolution and the compositional pathways during isothermal precipitation at 978 K (705 °C) in a model Ni-6.5Al-9.9Mo at.% superalloy. We demonstrate that phase separation occurs by concurrent coagulation and coalescence of the $\gamma'(L1_2)$-precipitates and with increasing aging time, the system enters a quasi-stationary coarsening regime. Phase-separation proceeds exceedingly fast at 978 K (705 °C), thus we fail to capture experimentally, via APT, the pure nucleation stage even when the shortest possible experimental aging time is employed. Atom-probe-tomography (APT) analyses permit the three temporal power-law exponents for coarsening to be determined experimentally and also via vacancy-mediated LKMC simulations.

In the phase separation of a solid-solution into two phases, an increase in the size scale of the second-phase precipitates reduces the total interfacial area of the precipitating phase, thus resulting potentially in morphological changes of the second phase [37]. The LSW model, derived from the work of Lifshitz and Slyozov [38] and Wagner [39], is limited to dilute binary alloys with several important and strong assumptions: (i) a vanishingly small volume fraction of the precipitating phase; (ii) precipitates have a spherical morphology; (iii) the precipitate diffusion fields do not overlap; (iv) the compositions of the two phases are equal to their equilibrium values, essentially at all times. Based on the above assumptions, the LSW model for a binary alloy yields the following relationships for the mean particle radius, $\langle R(t) \rangle$, the number of precipitates per unit volume, $N_v(t)$, and the matrix supersaturations,

$\Delta C_i^{\gamma}(t)$:

$$\langle R(t)\rangle^p - \langle R(t_0)\rangle^p = K(t-t_0) \tag{1}$$

$$N_v(t)^q = K_n(t-t_0) \tag{2}$$

$$\Delta C_i^{\gamma}(t) = \langle C_i^{\gamma,ff}(t)\rangle - C_i^{\gamma,eq}(\infty) = \kappa(t-t_0)^{-1/r} \tag{3}$$

where $K$, $K_n$ and $\kappa$ are the associated rate constants for $\langle R(t)\rangle$, $N_v(t)$ and $\Delta C_i^{\gamma}(t)$, respectively; $\langle R(t_0)\rangle$ is the mean precipitate-radius at the onset of quasi-stationary coarsening at $t_0$. The quantity $\Delta C_i^{\gamma}(t)$ is the difference between the concentrations in the far-field (*ff*) matrix, $\langle C_i^{\gamma,ff}(t)\rangle$, and the equilibrium matrix's solute-solubility, $C_i^{\gamma,eq}(\infty)$.

Following the work of LSW, Umantsev and Olson (UO) [40] removed the limitations of dilute-solution thermodynamics on multi-component coarsening; the UO model did not, however, permit the composition of the precipitates to evolve temporally, similiarly for the work of Morral and Purdy [41]. Kuehmann and Voorhees (KV) [42] considered isothermal quasi-stationary coarsening of ternary alloys; however, capillarity affects the precipitate's mean composition, such that both the matrix's and the precipitate compositions can deviate from their equilibrium thermodynamic values. In the KV model, only the diagonal terms in the diffusion matrix are considered. Based on the KV model, Philippe and Voorhees (PV) [43] developed a more general model of coarsening in multicomponent alloys, which is valid for non-ideal and non-dilute solutions, and which accounts for the off-diagonal terms in the diffusion tensor. Both the KV [42] and PV [43] models's temporal exponents for the power laws for the mean radius, number density of precipitates, and the supersaturations in the matrix are the same as for a binary alloy. From the PV model, the behavior of these quantities during coarsening depends on both thermodynamic and kinetic factors, through the Hessian of the matrix's free-energy and the mobility tensor. The mobility tenor and diffusion matrix strongly affect the length of a vector representing the deviation of the precipitate's compositions from their equilibrium values. In this research，the quantity $C_i^{\gamma,eq}(\infty)$ must be calculated or determined experimentally by using non-linear multivariate regression analyses because the phase diagram is unavailable for the Ni-Al-Mo ternary alloys at 978 K (705 °C).

## 2. Experimental, analytical and theoretical procedures

### 2.1 Materials and processes

High-purity constituent elements (99.97 wt.% Ni, 99.98 wt.% Al, and 99.99 wt.% Mo) were induction-melted and chill cast in a 19 mm diameter copper mold under an argon atmosphere. The overall composition of the Ni-Al-Mo alloy was determined utilizing inductively-coupled plasma (ICP) atomic-emission spectroscopy (AES) at NASA Glenn Research Center, which yielded Ni-6.50Al-9.90Mo at.%. Chemical homogeneity of the cast ingots was achieved by annealing at 1448 K (1175 °C) in the γ(f.c.c.)-phase field for 4 h, followed by water quenching to room temperature prior to being sectioned. The ingot sections were then aged at 978 K (705 °C) under flowing argon for aging times ranging from 1/6 to 1024 h, followed by water quenching into a salt-solution. APT nanotip specimens were prepared from each of the aged sections utilizing an electrochemical polishing technique. Rods ($0.2 \times 0.2 \times 10$ mm$^3$) were cut from the ingots and electrochemically polished in 10% perchloric acid in glacial acetic acid electrolyte at 10 to 15 Vdc and then in 2% perchloric acid in 2-butoxyethanol solution at 5 Vdc [44] at room temperature.

### 2.2 Analytical methods

### 2.2.1 Atom-probe tomography analyses

APT analyses, utilizing a pulsed picosecond ultraviolet (UV) laser (wavelength = 355 nm), were performed employing an evaporation rate of 0.04 ion pulse$^{-1}$, a specimen base temperature of 25.0±0.3 K, a laser pulse energy of 20-30 pJ pulse$^{-1}$, a pulse repetition rate of 200 kHz, and a gauge pressure of $<6.7 \times 10^{-8}$ Pa. APT data were visualized and analyzed utilizing the software package IVAS3.6.2 (Cameca, Madison, WI).

Within the reconstructed volumes, the γ(f.c.c.)/γ'(L1$_2$) heterophase-interface is delineated employing an Al isoconcentration surface [45]. The threshold value, $\eta$ (Al), of the isoconcentration surface is defined as the average value of the mean Al concentration values in the γ(f.c.c.)-phase and γ'(L1$_2$)-precipitate-phase away from the heterophase interfaces. These concentration values are referred to as plateau concentrations because a flat concentration profile is anticipated at distances away from the γ(f.c.c.)/γ'(L1$_2$)-heterophase-interface. The plateau concentrations of Al are determined by averaging the values that are part of the flat Al concentration-profile in a proximity histogram [46]. By utilizing the average value of the Al concentrations, a unique value of $\eta$(Al) is determined for each analysis, Table 1. The volume fraction of the $\gamma'(L1_2)$-precipitate-phase, $\phi(t)$, is defined as the ratio of the total number of atoms contained within the isoconcentration surfaces, delineating the $\gamma'(L1_2)$-precipitate-phase, to the total number of atoms collected in an analyzed volume. By using this direct counting method, the precipitate number density, $N_v(t)$, is given by the number of $\gamma'(L1_2)$-precipitates contained in the reconstructed 3D volume in accordance with the following conventions [18]: (a) a single-precipitate contained fully within the analyzed volume contributes one precipitate; (b) a single-precipitate contained partially within the analyzed volume contributes 0.5 precipitates; and (c) a coalesced pair of precipitates, contained fully within the analyzed volume, contributes two precipitates to the total $N_v(t)$ count. The same conventions are employed in determining the percentage of $\gamma'(L1_2)$-precipitates interconnected by necks, $f(t)$. The edge-to-edge inter-precipitate spacing, $\lambda_{edge-edge}$ (t), was calculated using an algorithm based on a Delaunay triangulation method [47].

Phase compositions in the far-field (*ff*) $\gamma$(f.c.c.)-matrix and in the $\gamma'(L1_2)$-precipitate cores are obtained from proximity histogram generated concentration profiles [46], which display the average concentrations in shells with a 0.25 nm thickness at a given specified distance from the $\gamma$(f.c.c.)/$\gamma'(L1_2)$-heterophase interfaces, Fig. 1. Obtaining the concentration profiles is accomplished by determining the number of atoms of each element within the shells associated with the plateau region of a specified concentration profile for a given phase, which is indicated by horizontal solid-lines, Fig. 1, and is delineated by a visual assessment of the error in the concentration measurements.

The atomic fraction of an element, $C_i$, in a given analysis volume is determined from APT with an uncertainty of $2\sigma_c$, where $\sigma_c$ is the standard deviation of an individual measurement within a population of $N$ atoms, $\sigma_c = \sqrt{\dfrac{C_i(1-C_i)}{N}}$. Standard errors, $2\sigma$, for $<R(t)>$, $N_v(t)$, $\phi(t)$, $f(t)$, and $\lambda_{\text{edge-edge}}$ (t), are calculated by utilizing counting statistics and reconstruction scaling errors [45], utilizing standard error propagation methods [48].

### 2.2.2 Morphology analysis

Discs with a 3 mm diameter were cut from the foil, mechanically ground to a thicknees of 150 μm, and electropolished employing a Struers double-jet electropolisher utilizing a solution of 8 vol.% perchloric acid and 14 vol.% 2-butoxyethanol in methanol at 243 K (-30 °C). This temperature was achieved by using a bath of dry ice in methanol. Scanning transmission electron microscopy (STEM) was performed utilizing an Hitachi HD 2300A STEM (Northwestern University) operating at 200 kV. Bright-field imaging techniques were performed employing a 001 superlattice reflection of the $\gamma'(L1_2)$-precipitate-phase, utilized for imaging precipitates.

### 2.2.3 Microhardness measurements

Vickers microhardnesses were measured by employing a Buehler Micromet instrument for samples polished to a root-mean-square roughness of 1 μm, for an applied load of 500 g, sustained for 10 s. The mean value of 15 independent measurements made on several grains was employed. Microhardness is an indirect measure of the precipitation sequence through strength changes, which depend on both $<R(t)>$ and $N_v(t)$.

### 2.2.4 Computational thermodynamics

Employing a Ni-based multicomponent ThermoCalc database developed by Saunders's, the commercial software package ThermoCalc [49] was utilized to calculate the values of the equilibrium volume fraction of the $\gamma'(L1_2)$-precipitate-phase , $\phi^{eq}$, and the composition for each solute species, $i$ , in both the $\gamma$(f.c.c.)-matrix- and $\gamma'(L1_2)$-precipitate-phases for the Ni-6.5Al-9.9Mo at.% alloy at a pressure of 1 atm and a temperature of 978 K (705 °C) [50]. There isn't an extant experimental Ni-Al-Mo phase diagram at 978 K (705 °C); therefore, the equilibrium phase boundaries of the Ni-Al-Mo system were determined by Thermo-Calc, using Saunders's database, and were superimposed on our Grand canonical Monte Carlo (GCMC) solvus curves (using the same interatomic interaction parameters as discussed below) in the phase diagram at 978 K (705 °C), Fig. 2. The ThermoCalc calculated $\gamma$ (f.c.c.)/ ($\gamma$(f.c.c.) plus $\gamma'(L1_2)$ solvus-curve is in good agreement with the GCMC results, whereas the $\gamma'(L1_2)$/ ($\gamma$(f.c.c.) plus $\gamma'(L1_2)$ solvus curves differ for the two techniques as displayed; the GCMC curve is significantly lower than the one predicted by ThermoCalc. The extrapolated equilibrium concentrations of the $\gamma'(L1_2)$- and $\gamma$(f.c.c.)-phases determined by APT experiments are in excellent agreement with our GCMC results.

### 2.2.5 Lattice kinetic Monte Carlo technique

The kinetics of the lattice kinetic Monte Carlo (LKMC) methodology are based on a thermally-activated diffusion process involving a monovacancy exchanging places with atoms (Ni or Al or Mo) in first nearest-neighbor (NN) sites [51]. The primitive atomic positions are an array of rhombohedral cells of the f.c.c. lattice and the volume of the simulation box is $N = L^3$, where L is 128 in all the regimes studied. One vacancy is introduced into the system: each lattice site is thus occupied by either one Ni or one Al or one Mo atom or by the monovacancy.

The exchange frequency, $W_{p,q}^{i,v}$, between an atom of type $i$ on site $p$ and a vacancy, $v$, on a NN site $q$ is given by:

$$W_{p,q}^{i,v} = v^i \exp\left(-\frac{E_{sp-p,q}^i - \sum_{broken\ bonds}\left(n_{ij}^k \varepsilon_{i-j}^k + \varepsilon_{v-i}^k\right)}{k_B T}\right);$$
(4)

where $v^i$ is the attempt frequency for the exchange, $E_{sp-p,q}^i$ is the binding energy of atom $i$ to the saddle point between sites $p$ and $q$, and $\varepsilon_{i-j}^k$ is the atomic interaction energy between atom i and j at the $k^{th}$ NN distance, $n_{ij}^k$ is the total number of i-j atomic pairs, $\varepsilon_{v-i}^k$ is the so-called ghost potential between *atom i* and a vacancy at a $k^{th}$ NN distance [52, 53]. The kinetic database requires additional parameters, specifically values for $v^i$ and the $E_{sp-p,q}^i$ terms, which are described in detail in reference [54] and are listed in Table 2.

A residence time algorithm (RTA) [55] was utilized. A configuration remains unchanged for certain physical time, which is related to the jump frequency of each first NN atom surrounding a vacancy. The time for each MC step is given by:

$$t = (\Sigma W^i)^{-1}$$
(5)

The vacancy-jump direction according to a probability for a jump frequency to each first NN atom at each MC step, must satisfy the following condition:

$$\sum_{i=1}^{j-1} W^i \leq \frac{\xi}{t} < \sum_{i=1}^{j} W^i \qquad (6)$$

where $\xi$ is a random number between 0 and 1. The effect of ordering on diffusion is automatically included in the LKMC simulations and the hence the γ'(L1$_2$)-precipitates exhibit order, which is time dependent [28].

### 2.2.6 The thermodynamic interatomic pair potentials

We use a cluster expansion of the cohesive energy of Ni(Al, Mo) supercells to determine values of $\varepsilon_{i-j}^k$ and $\varepsilon_{v-i}^k$ from first-principles calculations. The calculations employed the plane-wave pseudo-potential total energy method, utilizing the generalized gradient approximation [56, 57], as implemented in the Vienna *ab initio* simulation package (VASP) [58-62], employing the projector augmented-wave potentials [63]. A plane-wave cutoff energy of 300 eV and 8×8×8 Monkhorst-Pack *k*-point grids were utilized. In this study, the long-distance homo-atomic interactions are obtained using the Chen-Möbius inversion-lattice technique [64]. The interaction parameters are listed in Table 3.

## 3. Results

### 3.1 Morphological development of γ'(L1$_2$)-precipitates

The full nanostructural evolution of γ'(L1$_2$)-precipitates was studied from 1/6 to 1024 h at 978 K (705 °C) for the 3D-APT reconstruction experiments. The temporal evolution of the morphology of the alloy is displayed in Fig. 3, which is compared with that of a Ni-5.2Al-14.2Cr at.% alloy [17, 21]. In parallel, LKMC simulations of the same alloys were performed through 256 hours at the same temperature. Qualitatively, the γ'(L1$_2$)-precipitates grow and coarsen, as demonstrated by the increase in $<R(t)>$ and a concomitant decrease in $N_v$(t) with increasing aging time from 1/6 h to 16 h and the morphology of the γ'(L1$_2$)-precipitates is an admixture of individual spheroidal γ'(L1$_2$)-precipitates and γ'(L1$_2$)-precipitates interconnected by necks, which suggests the coagulation and coalescence mechanism dominates during this nucleation and growth stage. In LKMC simulations, for time <5 min, $<R(t)>$ is a constant and $N_v$(t) increases linearly, which demonstrates clearly that this is the nucleation stage.

The formation of necks interconnecting pairs of γ'(L1$_2$)-precipitates is evidence for the dominating coagulation and coalescence mechanism in three different Ni-Al-Cr alloys [17, 21]. Our studies on Ni-Al-Cr alloys indicate that there are two precipitation growth and coarsening mechanisms that exist in solute-concentrated alloys: (1) the evaporation-condensation mechanism (the big eat the small); and (2) the coagulation-coalescence mechanism. Typically, the coagulation-coalescence mechanism occurs at aging times, where the edge-to-edge distance between precipitates is small and $N_v$(t) is large, which results in a large fraction of the precipitates being interconnected by γ'(L1$_2$)-precipitates [19, 40, 41]. The evaporation - condensation mechanism occurs when $N_v$(t) is small and the edge-to-edge distance between precipitates is large , which happens in the pure nucleation regime and at long aging times in the pure coarsening regime. We observe similar coagulation - coalescence behavior among precipitates in this Ni-Al-Mo alloy, Fig. 4, which displays a large fraction of interconnected γ'(L1$_2$)- precipitates as early as 1/6 h at 978 K (705 ºC) with $f$ = 21.02 ± 1.02 %. The percentage of coagulating and coalescing $\gamma'(L1_2)$-precipitates decreases continuously with increasing aging time. Interconnected $\gamma'(L1_2)$-precipitates are still detected when the aging time is ≤ 16 h. When the aging time is >16 h, there are no longer $\gamma'(L1_2)$-precipitates interconnected by necks because the evaporation-condensation occurs for a second time. In contrast, the edge-to-edge inter-precipitate distance, $\lambda_{edge-edge}(t)$, increases monotonically because $N_v(t)$ is continuously decreasing.

Fig. 5(a) demonstrates that the $\gamma'(L1_2)$-precipitates are interconnected by necks at 1/6 h with $f$ = 21.02 ± 1.02 %. An examination of the atoms in the $\gamma'(L1_2)$-precipitate-pairs, Fig. 5(b), demonstrate that an interplanar distance of 0.356 nm is present, which corresponds to the {100} lattice planes of the $\gamma'(L1_2)$-structure. The atomic planes extend through the necks, implying that they have the same crystallographic $L1_2$-structure. The LKMC simulations demonstrate that the coagulation-coalescence mechanism occurs at less than 5 min with $f$ = 22 ± 0.6 % because the $\gamma'(L1_2)$-precipitates are already interconnected by necks at 5 min, Fig. 5(c).

### 3.2 Temporal evolution of the nanostructural properties of $\gamma'(L1_2)$-precipitates

In scanning transmission electron microscope (STEM) images there are problems concerning the overlap of precipitates and truncation effects due to the finite thickness of TEM foils. Therefore, we studied the temporal evolution of the following nanostructural properties of $\gamma'(L1_2)$-precipitates using APT: $<R(t)>$, $N_v(t)$, $\phi(t)$, and $f(t)$, which can be determined more accurately from APT results than from STEM images.

The values of these nanostructural properties, $<R(t)>$, $N_v$(t), $\phi(t)$, and $f(t)$, are summarized in Table 1. These properties are based on the number of analyzed precipitates, $N_{ppt}$, listed in Table 1. The temporal evolution of $\phi(t)$, $N_v$(t) and $<R(t)>$ as a function of aging time, from both APT experiments and LKMC simulation results, are displayed in Fig. 6.

From LKMC simulation results we find that a pure nucleation regime occurs within a 5-min aging time, due to fast atomic diffusion kinetics at 978 K (705 °C), whereas we cannot find a pure nucleation regime in the APT samples because the shortest aging time is 10 min. The coagulation-coalescence mechanism of γ′(L1$_2$)-precipitates occurs at an early aging time of less than 10 min. We have already discussed in detail what occurs when two neighboring γ′(L1$_2$)-precipitates merge to form a single γ′(L1$_2$)-precipitate [17, 28, 35, 36]. The temporal evolution of γ′(L1$_2$)-precipitates in the Ni-6.5Al-9.9Mo at.% alloy aged at 978 K (705 °C) is complex. Four regimes are determined by LKMC simulations, however, the last three are identified from the APT results: (1) nucleation (less than 5 min, determined by LKMC); (2) concomitant precipitate nucleation and growth prior to 1/4 h; (3) concurrent coagulation and coalescence from 1/2 h to 16 h; and (4) quasi-stationary coarsening of γ′(L1$_2$)-precipitates from 16 h to 1024 h, where no interconnected γ′(L1$_2$)-precipitates are detected utilizing APT. We discuss regimes (2) to (4), as identified by APT, below.

### 3.2.1. Nucleation and growth of the γ′(L1$_2$)-precipitates (t ≤ 1/4 h)

In this regime, $t \leq 1/4\ h$, a high number density, $N_v$(t), (1.88±0.04)×10$^{24}$ m$^{-3}$, of γ′(L1$_2$)-precipitates is detected after 1/6 h. The corresponding values of $\phi(t)$ and $<R(t)>$ are 7.76 ± 0.18 % and 2.43 ± 0.78 nm, respectively. At an aging time between 1/6 h to 1/4 h, the Ni-6.5Al-9.9Mo at.% model superalloy is in a regime of concomitant precipitate nucleation and growth, which results in a steadily increasing value of $\phi(t)$, from 7.76±0.18% to 8.25±0.24%, and $<R(t)>$ is continuously increasing. The nucleation stage is only observed in the LKMC simulations for time <5 min, where $<R(t)>$ is a constant and $N_v(t)$ increases linearly.

### 3.2.2. Coagulation and coalescence of the γ'(L1₂)-precipitates (1/4 < t ≤ 16h)

Coagulation and coalescence of γ'(L1$_2$)-precipitates occur concurrently from 1/4 h to 16 h as $N_v(t)$ decreases from $(1.78±0.05)×10^{24}$ to $(0.97±0.07)×10^{23}$ m$^{-3}$, and $<R(t)>$ increases monotonically from 2.51±0.81 to 8.13±2.56 nm. Simultaneously, $\phi(t)$ increases to 16.68±1.22%, which agrees with the calculated equilibrium volume fraction value, $\phi_{calc}^{eq}$, of 16.86%, obtained employing Thermo-Calc [49] employing Saunders's thermodynamic database for nickel-based alloys [50]. A nonlinear linear multivariate regression analysis of $<R(t)>$ for 1/4 < $t$ ≤ 16h yields a coarsening rate constant, $K_{KV}$, of $(6.06±0.26)×10^{-30}$ m$^3$s$^{-1}$, with a temporal power-law exponent, 0.35±0.01, which is in agreement with the 1/3 value for coarsening [38, 39, 42]. During this region, $N_v(t)$ decreases with a temporal power-law exponent, -0.87±0.05, which is less than the -1 value predicted for coarsening [38, 39, 42]. As $<R(t)>$ increases with aging time, the coarsening relationship demonstrates that $N_v(t)$ does not decay as $t^{-1}$ when $\phi$ is less than its equilibrium value (that is, when phase separation is incomplete). Fig. 4 demonstrates that the percentage of γ'(L1$_2$)-precipitates interconnected by necks, $f(t)$, decreases toward zero with increasing time, and no interconnected $\gamma'(L1_2)$-precipitates with necks are detected after being aged at 978 K (705 °C) for > 16 h, demonstrating that the coarsening mechanism transforms from the coagulation-coalescence mechanism of $\gamma'(L1_2)$-precipitates to the evaporation-condensation mechanism (the big eat the small) [28].

### 3.2.3. Quasi-stationary coarsening of $\gamma'(L1_2)$-precipitates ($t > 16$ h)

During quasi-stationary coarsening (aging times from 16 to 1024 h), the value of $N_v(t)$ decreases sharply with a temporal dependence of $t^{-0.95\pm0.02}$, with a value of $(1.02\pm0.07)\times10^{23}$ m$^{-3}$ at 16 h to $(1.14\pm0.35)\times10^{21}$ m$^{-3}$ at 1024 h. And concomitantly the temporal dependence of $<R(t)>$ is $0.344\pm0.012$. Thus, values of the temporal dependencies for $N_v(t)$ and $<R(t)>$ are in good agreement with the predicted values of -1 and 1/3, respectively, from the LSW [38, 39], UO [40], KV[42] and PV [43] models for coarsening.

### 3.3 The temporal evolution of the $\gamma'(L1_2)$-precipitate size distributions (PSDs) and edge-to-edge inter-precipitate distances

The temporal evolution of the $\gamma'(L1_2)$-precipitate size distributions (PSDs) as a function of the normalized quantity, $R/<R(t)>$, on the abscissa is plotted in Fig. 7 for the Ni-Al-Mo alloy aged at 978 K (705 °C), which are compared with the Akaiwa-Voorhees (AV) PSD for the $\phi^{eq} = 20\%$ [65]. A bin size of 0.2 nm is employed for $R/<R(t)>$. For the ordinate axis, the number of $\gamma'(L1_2)$-precipitates in each bin is divided by two quantities: (1) the total number of enclosed $\gamma'(L1_2)$-precipitates; and (2) the bin size (0.2 nm). So that the PSDs for different aging times are directly comparable, even though $<R(t)>$ is different for each aging time. The vacancy-mediated LKMC results are not used to generate PSDs because the analyzed volume is too small to yield satisfactory statistics. The stationary predictions are for volume fractions that are asymptotically approaching the measured $\phi^{eq}$ value of 16.86% in our Ni-Al-Mo alloy. For the specimens aged for 1/4 to 1 h, the PSDs are broader than the stationary PSD, but the maximum appears at a smaller $R/<R(t)>$ value of 1.0 according to the AV model. With further aging to 16 h, the maximum value shifts to the $R/<R(t)>$ value predicted by the AV model, 1.16 [24]. In the final aging state, $t = 256$ h, the PSD becomes narrower as the height of the maximum increases and remains at the $R/<R(t)>$ value of 1.16, as predicted by the AV model [65]. Similar results were measured employing small-angle scattering and TEM for Ni-8.8Al- 9.6Mo and Ni-9.5Al-5.4Mo at.% alloys aged at 923 K (655 °C) [35, 36].

The frequency distributions of $\lambda_{edge-edge}$ (divided into a constant number of intervals over the full range of values) as a function of aging time are displayed in Fig. 8 and summarized in Table 1. The mean value of $\lambda_{edge-edge}$ is 6.76±3.24 nm at t = 1/6 h and increases monotonically to 45.47±39.07 nm at t = 256 h.

### 3.4 Temporal evolution of the compositions of the $\gamma$(f.c.c.)-matrix and $\gamma'(L1_2)$-precipitates

The compositional trajectories, as measured by APT, of both the $\gamma$(f.c.c.)-matrix and the γ'(L1$_2$)-precipitate-phases are displayed in Figure 2; the black arrows indicate the direction of increasing time. The compositional trajectory of the γ'(L1$_2$)-precipitate-phase commences inside the γ'(L1$_2$)-phase-field and it evolves temporally toward the solvus curve between the γ'(L1$_2$)-phase- and the (γ(f.c.c.) plus γ'(L1$_2$))-phase-fields. The compositional trajectory of the γ(f.c.c.)-matrix commences at the mean composition of the Ni-Al-Mo alloy and evolves toward the solvus curve between the (γ(f.c.c.) plus γ'(L1$_2$)-phase-fields) and the γ(f.c.c.)-phase-field. The extrapolated equilibrium composition of the γ(f.c.c.)-phase is close to that calculated value using GCMC simulations. Table 5 displays the compositions of both the γ(f.c.c.)-matrix and the γ'(L1$_2$)-precipitate-phase as they evolve temporally. The γ(f.c.c.)-phase becomes enriched in Ni and Mo and depleted in Al, whereas the γ'(L1$_2$)-precipitate-phase becomes enriched in Ni and depleted in both Mo and Al. The γ'(L1$_2$)-precipitate-phase contains Al (17.81±0.10 at.%) and Mo (7.33±0.18 at.%) at $t$ = 1/6 h, which decrease continuously to Al (16.91±0.08 at.%) and Mo (6.51±0.05 at.%) at $t$ =1024 h, indicating that the γ'(L1$_2$)-precipitates are initially supersaturated in Al and Mo. With increasing aging time, the far-field concentrations of Al in the γ(f.c.c.)-matrix decrease, whereas the Mo concentrations increase (Table 5), which is characteristic of decreasing supersaturations as Al partitions to the γ'(L1$_2$)-precipitates and Mo to the γ(f.c.c.)-matrix.

As $t$ >16 h, coagulation and coalescence of γ'(L1$_2$)-precipitates aren't detected. As phase separation progresses and the magnitude of the values of $\Delta C_i^{\gamma}(t)$ decrease asymptotically toward a value of zero as the equilibrium γ(f.c.c.)-matrix and γ'(L1$_2$)- precipitate phase compositions are approached, implying that the system is approaching a stationary-state [66]. The equilibrium compositions of the γ(f.c.c.) -matrix and γ'(L1$_2$)-precipitate are extrapolated by using a nonlinear multivariate regression analysis to fit the measured concentrations from the quasi-stationary coarsening regime to Eq. (3) for aging times beyond 16 h. The equilibrium compositions of the γ(f.c.c.)-matrix is estimated to be Ni (85.21±0.01 at.%), Al (4.12±0.02 at.%), and Mo (10.67±0.01 at.%). Similarly, the equilibrium γ'(L1$_2$) -precipitate-phase has a composition of Ni (76.64±0.12 at.%), Al (16.89±0.06 at.%), and Mo (6.47±0.06 at.%) at infinite time.

The partitioning ratio, $\kappa_i$, is defined as the concentration of element $i$ in the γ'(L1$_2$)-phase, divided by its concentration in the γ(f.c.c.)-phase [22], $\kappa_i = C_i^{\gamma'} / C_i^{\gamma}$, where $i$ = Ni, Al or Mo: the quantity $\kappa_i$ is utilized to quantify elemental partitioning behavior. Fig. 9 demonstrates that our Ni-Al-Mo alloy, from for both the APT and LKMC results, exhibit partitioning of Al to the γ'(L1$_2$)-precipitates and of Mo to the γ(f.c.c.)-matrix, where the values of $\kappa_{Al}$ and $\kappa_{Mo}$ are time-dependent. Nickel partitions, however, to the γ(f.c.c.)-matrix with a $\kappa_i$ of ~0.9, and it is only slightly time-dependent. After 1024 h of aging, $\kappa_{Mo}$ decreases to 0.61±0.01, indicating that Mo partitions preferentially to the γ(f.c.c.)-matrix. The reason for this behavior is that the calculated site substitutional energy of Mo in the γ(f.c.c.)-matrix, 0.085 eV atom$^{-1}$, is smaller than that in γ'(L1$_2$)-precipitate, 0.792 eV atom$^{-1}$. The details of this calculation are elaborated on in our prior study, which combined APT and first-principles calculations [67].

The values of the γ(f.c.c.)-matrix and γ'(L1$_2$)-precipitate-phase supersaturations of our Ni-Al-Mo alloy are calculated on the basis of the equilibrium phase - compositions from the APT data. The initial supersaturation values are estimated to be 0.89±0.02 at.% Al and 0.43±0.03 at.% Mo in the γ(f.c.c.)-matrix, and 0.93±0.08 at.% Al and 0.85±0.09 at.% Mo in the γ'(L1$_2$)-precipitate. The temporal evolution of the Al and Mo γ(f.c.c.)-matrix supersaturation values are presented in Fig. 10(a), where the $\Delta C_{Al}^{\gamma}(t)$ values are positive, whereas the $\Delta C_{Mo}^{\gamma}(t)$ and $\Delta C_{Ni}^{\gamma}(t)$ values are negative. During quasi-stationary coarsening, $t$ >16 h, the diminution of the $\Delta C_i^{\gamma}(t)$ values follow the $t^{-1/3}$ prediction of the KV and PV models [42, 43]. Using a non-linear multivariate regression analysis, Fig. 10(a) displays the temporal dependencies of $t^{0.312±0.012}$ for $\Delta C_{Al}^{\gamma}(t)$, $t^{-0.314±0.025}$ for $\Delta C_{Mo}^{\gamma}(t)$, and $t^{-0.314±0.008}$ for $\Delta C_{Ni}^{\gamma}(t)$. Fig. 10(b) displays the supersaturation values of the γ'(L1$_2$)-precipitate-phase, $\Delta C_i^{\gamma'}(t)$. The quantities $\Delta C_{Al}^{\gamma'}(t)$, $\Delta C_{Mo}^{\gamma'}(t)$, and $\Delta C_{Ni}^{\gamma'}(t)$ exhibit temporal dependencies of $t^{-0.282±0.024}$, $t^{-0.313±0.015}$, and $t^{-0.343±0.088}$, respectively, which are also close to the predicted value of $t^{-1/3}$ of the three coarsening models, LSW, UO, KV and PV.

### 3.5. Temporal evolution of the interfacial concentration width

The interfacial compositional width, $\delta_i(t)$, is determined by measuring the horizontal distance between the 10 and 90 % values for the plateaus of the concentration profiles of the γ-(f.c.c.) matrix and γ'(L1$_2$)-precipitates. The $\delta(t)$ values of Ni, Al and Mo are determined from the measured elemental concentration profiles by a spline-function fitting procedure [29] as opposed to assuming a function for the concentration profiles; this is important as the concentration profiles are not symmetric. Next, each measured $\delta_i(t)$ value is normalized by its corresponding $<R(t)>$-value, Table 1; the measured $\delta_i(t)$ values and the resulting normalized interfacial concentration widths, $\delta_i(t)/<R(t)>$, are listed in Table 5. As demonstrated in Fig. 11, both the $\delta_i(t)$ and $\delta_i(t)/<R(t)>$ values decrease with increasing $<R(t)>$ values and aging times, ranging from 0.167 to 1024 h. And more specifically the $\delta_i(t)/<R(t)>$ values, as defined by the Ni, Al, and Mo concentration profiles, decrease from 1.13±0.36 for an aging time of 0.167 h to 0.06±0.03 at 1024 h, and from 1.37±0.44 to 0.06±0.03, from 1.31±0.42 to 0.08±0.04, respectively, for the Ni, Al and Mo concentration profiles. The data in Fig.11(c) indicate that the values of $\delta_i(t)$ are decreasing continuously as $<R(t)>$ is increasing, and the values of $\delta_i(t)$ are asymptotically approaching a constant value at the longest aging time, 1024 h, which are definitely not equal to zero. The data in Fig.11 (c) fit a $<R(t)>^n$-type relationship utilizing nonlinear multivariate regression analyses, where $n_{Al}$= -2.06±0.80, $n_{Mo}$= -1.71±0.35 and $n_{Ni}$= -0.55±0.26. While, the interfacial compositional widths at infinite ageing time, $\delta_i(t)$, are estimated to be 1.89±0.22, 2.09±0.12 and 2.64±0.03 nm for Ni, Al and Mo, respectively. Our LKMC results agree with the 3-D APT results, but they slightly under estimate the value of $\delta(t)$.

## 4. Discussion

### 4.1 Temporal evolution of the Vickers microhardness values

The temporal evolution of the nucleation, growth, and coarsening regimes for our Ni-Al-Mo ternary alloy are studied by correlative APT experiments and LKMC simulations. The original assumptions of the Lifschitz-Slyozov-Wagner (LSW) analyses of Ostwald ripening are that the γ'(L1$_2$)-precipitates are coherent, spheroidal, with relatively stress-free matrix/precipitate heterophase interfaces. During phase separation from a single-phase solid-solution, microhardness measurements provide an indirect measurement of the precipitation sequence through strength changes, which is a function of the precipitates's $<R(t)>$ and $N_v(t)$ values. Fig. 12 displays the evolution of the microhardness as a function of aging time. Prior to 16 h, the microhardness increases gradually from 1995±69 MPa to 2413±111 MPa. After 16 h, it increases rapidly, reaching a peak value of 3681±127 MPa at an aging time of 256 h. After the peak value, the microhardness decreases due to the phenomenon of over-aging. Scanning transmission electron microscope (STEM) images demonstrate that the γ'(L1$_2$)-precipitates in a specimen aged at 978 K (705 ºC) for 1024 h remain spheroidal and are homogeneously distributed, Fig. 13, which implies that the γ'(L1$_2$)-precipitates maintain coherency with the γ(f.c.c.)-matrix and the morphology of the precipitate is determined essentially by the γ(f.c.c.)/ γ'(L1$_2$) interfacial free energy, which is not a function of {hkl}.

### 4.2 Morphological and nanostructural evolution of the γ'(L1$_2$) precipitates

At 978 K (705 ºC) phase-separation proceeds extremely fast during early aging times, and as a result using APT, we miss the transition between the nucleation and the growth stages. We observe significant γ'(L1$_2$)-precipitate growth and the competitive coarsening stage at the shortest possible aging time, 10 min. Although often treated as distinct processes, nucleation, growth, and coarsening may occur concurrently as precipitation proceeds [68, 69]. Generally, there are two precipitation growth and coarsening mechanisms that can exist in alloys with high solute-concentrations due to the coupling of diffusional fluxes: (1) the classical evaporation-condensation mechanism, which is implicit in the LSW model; and (2) the coagulation-coalescence mechanism, which is not so common. In the present research, the coagulation-coalescence mechanism occurs at aging times of less than 16 h, where the edge-to-edge distance between $\gamma'(L1_2)$-precipitates is small, $<20.45\pm15.57$ nm, and the precipitate number density is high, $>0.968\pm0.07\times10^{23}$ m$^{-3}$. After an aging time of 16 h the evaporation-condensation mechanism becomes dominant for a second time. Three regimes are studied experimentally, via APT, and are identified during the temporal evolution of this alloy: (1) concomitant $\gamma'(L1_2)$-precipitate nucleation and growth (t = 1/6 to 1/4 h); (2) concurrent coagulation and coalescence of $\gamma'(L1_2)$-precipitates (t =1/4 to 16 h); and (3) quasi-stationary coarsening of $\gamma'(L1_2)$-precipitates (t = 16 to 1024 h).

Coalescence, in Ni-based superalloys, between coherent $\gamma'(L1_2)$-precipitates during precipitation, without an external stress, has rarely been reported and is only observed in alloys with a relatively high volume fraction of $\gamma'(L1_2)$-precipitates, $\phi^{eq}>35\%$, during the intermediate to late stages of coarsening [8, 35, 36, 70, 71]. Sequeira et al. [35, 36] suggested that coalescence is possibly driven by the removal of the elastically strained matrix material between the $\gamma'(L1_2)$-precipitates in Ni-Al-Mo alloys. In a similar alloy, Ni-6.5Al-9.8Mo at.%, aged at 1023 K (750 ºC), with near-zero lattice parameter mismatch values were studied for the entire range of phase separation [33]. Based on a uniform distribution of spheroidal $\gamma'(L1_2)$ precipitates, we think that the $\gamma'(L1_2)$-precipitates probably also have a small lattice parameter misfit with respect to the $\gamma$(f.c.c.)-matrix in our Ni-Al-Mo alloy. As presented in Section 3.2, the size of $\gamma'(L1_2)$-precipitates undergoing coagulation and coalescence are small, $<R(t = 16\ h)> \leq 8.13\pm2.56$ nm, and the lattice parameter misfit between $\gamma'$ ($L1_2$)-precipitates and $\gamma$(f.c.c.)-matrix is very small, the coherency strains between the precipitates and matrix are negligible. Therefore, elastic strain energy effects are less significant for coagulation and coalescence in our experiments as the $\gamma'$ ($L1_2$)-precipitates remain spheroidal during coarsening.

Sudbrack et al.[17] presented the first direct-space evidence for a coagulation and coalescence mechanism for coarsening in a Ni-5.2Al-14.2Cr at.% alloy aged at 873 K (600 °C), using a 3-D atom-probe field-ion microscope (APFIM), during early-stage phase separation with a small volume fraction of $\gamma'(L1_2)$-precipitates, $\phi$(t =1/4 h) = 0.55±0.06%, with 9±3% of the $\gamma'(L1_2)$-precipitates identified as being interconnected by necks. Correlating the APT results, LKMC simulations and diffusion theory, Mao et al. [28] explained the origin of the necks between adjacent $\gamma'(L1_2)$-precipitates, which occurs abundantly in the early stages of phase separation. It results from the overlap of the diffusion concentration profiles surrounding the $\gamma'(L1_2)$-precipitates. Therefore, the coagulation and coalescence mechanism depends strongly on the mean edge-to-edge inter-precipitate distance, $\lambda_{edge-edge}(t)$, values among neighboring $\gamma'(L1_2)$-precipitates; only $\gamma'(L1_2)$-precipitates whose edges are within a critical distance may coagulate and then coalesce. The critical distance can be determined from the concentration profiles crossing the γ(f.c.c.)/γ'(L1$_2$)-heterophase interface, which we take to be the interfacial width, $\delta_{Al}(t)$, determined using a spline fit of the concentration profile of Al [29], Table 5.

For our Ni-Al-Mo alloy after being aged at 978 K (705 °C) for >16 h, no evidence for a coagulation and coalescence mechanism is detected and the volume fraction of γ'(L1$_2$)-precipitates, $\phi(t)$, is approaching the equilibrium volume fraction value, indicating that phase-separation proceeds into the quasi-stationary coarsening regime. The temporal dependencies of $N_v$(t) and $<R(t)>$, -0.95±0.02 and 0.344 ± 0.012, respectively, agree with the predicted values of -1 and 1/3 for coarsening models [40, 42]. The evaporation-condensation mechanism is therefore the underlying mechanism for coarsening (Ostwald ripening) in this regime, which assumes that single atoms evaporate from smaller precipitates and diffuse through the matrix via a random walk process and eventually condense on larger precipitates [68, 69]. This mechanism obtains after 16 h of aging at 978 K (705 °C) for our Ni-Al-Mo alloy.

This value of $q = -1$ for $N_v(t)$, Eq. 2, differs from a mathematical equation posited by Ardell [72] and Xiao and Haasen [73], who utilized the relationship, $N_v(t) = c_1 t^{-1} - c_2 t^{-4/3}$ where $c_1$ and $c_2$ are given in references [72, 73]. This mathematical relationship is not generally correct. In references [74-76], Marqusee and Ross demonstrate rigorously that as time approaches infinity $<R(t)>$ becomes proportional to $t^{1/3}$ and $N_v(t)$ is proportional to $t^{-1}$. If one adds mathematical corrections terms to these laws, as hypothesized by Ardell [72] and Xiao and Haasen

[73], then it is necessary to add higher order terms to all the pertinent physical quantities,
$< R(t) >$, $N_v(t)$, $\Delta C_i^{\gamma}(t)$, and the PSD, which are then no longer unique.

### 4.3 Temporal evolution of the compositional interfacial widths, $\delta(t)$, between the γ (f.c.c.)- and γ' (L1₂)-phases

We have find that the average value of interfacial width, $\delta(t)$, decreases with increasing aging time, Fig. 11, which is consistent with the Cahn-Hilliard [77, 78] and Martin models [79], and with our prior 3-D APT studies [22, 80, 81], although it is never equal to zero, that is, an atomically sharp interface is never obtained. In parallel with the APT experiments, we find that the $\delta(t)$ values for our Ni-Al-Mo alloy decreases with increasing aging time by performing monovacancy-mediated LKMC simulations, which include monovacancy-solute binding energies to 4th NN distances and atom-atom interactions that also extend to 4th NN distances. The $\delta(t)$ values are determined by atomistic interactions that include non-zero monovacancy–solute binding energies. The atom-atom and monovacancy-solute binding energies are not functions of aging time or <R(t)>. Therefore, the compositional thickness of the transition layer between the γ(f.c.c)- and γ'(L1₂)-precipitate-phases is affected by the details of the diffusion mechanism; indeed, decreasing the range of monovacancy-solute interactions, modifies the correlations among the solute atoms, Al and Mo, and solvent, Ni, fluxes. The interaction distance over which atom-atom and monovacancy-solute interactions occur is a constant for a given alloy at a specified temperature. Therefore, the ratio of the interaction distance and <R(t)> of precipitates must decrease with increasing aging time, because the interaction distance is a constant and $<R(t)>$ is continuously increasing, which is the situation for an alloy becoming at least quasi-stationary. Our results are at variance with a mathematical ansatz posited by Ardell [82] for his trans-interface diffusion-controlled coarsening (TIDC) model, which conjectures that $\delta(t)$ increases with increasing $<R(t)>$ [82] and this is at variance with our APT and LKMC results.

## 5. Summary and Conclusions

We investigated the morphological, nanostructural, and phase compositional evolution of a model Ni-6.5Al-9.9Mo at. % superalloy during isothermal phase separation at 978 K (705 °C) for aging times ranging from 1/6 to 1024 h utilizing atom-probe tomography, vacancy-mediated lattice kinetic Monte Carlo simulations, microhardness measurements and transmission electron microscopy, and we arrive at the following conclusions:

- On the basis of the temporal evolution of the morphologies, the number density of $\gamma'(L1_2)$-precipitates, $N_v(t)$, the mean precipitate radius, $<R(t)>$, and the volume fraction of $\gamma'(L1_2)$-precipitate, $\phi(t)$, four phase-separating regimes of the model Ni-6.5Al-9.9Mo at.% alloy are identified (three are identified by APT): (1) nucleation stage (t < 5 min, LKMC simulations); (2) concurrent precipitate nucleation and growth (t <1/4 h); (3) concurrent coagulation and coalescence (t = 1/4 to 16 h); and (4) quasi-stationary coarsening of $\gamma'(L1_2)$-precipitates (t = 16 to 1024 h). These four regimes permit the full range of phase separation processes to be investigated in great detail.

- As a result of a near-zero lattice parameter misfit between the $\gamma$(f.c.c.)-matrix and

γ'(L1$_2$)-precipitate-phases, a uniform distribution of spheroidal γ'(L1$_2$)-precipitates is retained through t = 1024 h. Coagulation and coalescence of γ'(L1$_2$)-precipitates are observed for aging times of 1/6 to 16 h as a result of overlapping non-equilibrium concentration profiles associated with adjacent γ'(L1$_2$)-precipitates. The fraction of the γ'(L1$_2$)-precipitates interconnected by necks, *f(t),* decreases monotonically to undetectable throughout the coagulation and coalescence stage (t = 1/6 to 16 h).

- In the quasi-stationary coarsening regime (t >16 h), the temporal dependence of $N_v(t)$, -0.95±0.02, is determined experimentally, whereas it is 0.344 ± 0.012 for $<R(t)>$, which correspond to the time law predictions of the LSW [38, 39], Umantsev-Olson [40], Kuehmann-Voorhees [42] and Philippe-Voorhees [43] models for coarsening of ternary alloys.

- Our determination of the temporal evolution of the number density, $N_v(t)$, demonstrates that the temporal exponent is -1. This value of $q$ = -1 for $N_v(t)$, Eq. 2, differs from a mathematical equation posited by Ardell [72] and Xiao and Haasen [73], who utilized the relationship $N_v(t) = c_1 t^{-1} - c_2 t^{-4/3}$, where $c_1$ and $c_2$ are given in [72, 73]. This relationship is not generally correct, Marqusee and Ross [74-76] for $N_v(t)$.

- Compared with the stationary predictions for stress-free coarsening at non-negligible volume fractions in the Akiwa-Voorhees (AV) model [65], the precipitate size distributions (PSDs) are initially skewed toward smaller radii. With increasing aging time, the PSDs approach the predicted AV distribution and are in near-agreement for an aging time of 256 h.

- Extrapolations from the γ(f.c.c.)-matrix far-field (*ff*) and γ'(L1$_2$)-precipitate core

concentrations to infinite time permits the determination of the solute-solubility in the γ(f.c.c.)-matrix phase of Ni-6.5Al-9.9Mo at.% at 978 K (705 °C) to be Ni (85.21±0.01 at.%), Al (4.12±0.02 at.%), and Mo (10.67±0.01 at.%), while the equilibrium γ'(L1$_2$)-precipitate composition is Ni (76.64±0.12 at.%), Al (16.89±0.06 at.%), and Mo (6.47±0.06 at.%).

- In the quasi-stationary regime (t> 16 h), the temporal exponents of the Ni, Al and Mo supersaturations in the matrix are -0.314±0.008, -0.312±0.012 and -0.314±0.025, respectively, while in the γ'(L1$_2$)-precipitates they are -0.343±0.088, -0.282±0.024 and -0.313±0.015, respectively, in reasonable agreement with the multi-component coarsening model's prediction of -1/3.

- In our Ni-Al-Mo alloy aged at 978 K (705 °C), Al partitions strongly to the γ'-(L1$_2$)-precipitate-phase, $\kappa_{Al}$ = 4.06±0.04 at t = 1024 h, while Mo and Ni partition to the γ(f.c.c.)-matrix with $\kappa_i$ values of 0.61±0.01 and 0.90±0.01 also at t = 1024 h, respectively.

- Both the interfacial concentration width, $\delta_i(t)$, and $\delta_i(t)/<R(t)>$ values decrease with increasing $<R(t)>$ values for aging times from 0.167 to 1024 h, but it never reaches zero. The vacancy-mediated LKMC results agree with the trend of the 3-D APT results, but they slightly under estimate the value of $\delta(t)$. The $\delta_{Ni}(t)/<R(t)>$ value decreases from 1.13±0.36 at 0.167 h to 0.06±0.03 at 1024 h, while $\delta_{Al}(t)/<R(t)>$ decreases from 1.37±0.44 to 0.06±0.03 and $\delta_{Mo}(t)/<R(t)>$ decreases from 1.31±0.42 to 0.08±0.04. The interfacial compositional widths at infinite ageing time, $\delta_i(t)$, are estimated to be 1.89±0.22 nm, 2.09±0.12 nm and 2.64±0.03 nm for Ni, Al and Mo, respectively. These results are at variance with Ardell's trans-interface-diffusion-coarsening (TIDC), which posited that the interfacial width increases with increasing aging time and increasing $<R(t)>$ [82], which are completely inconsistent with our APT and LKMC results.

**Acknowledgements**

This research was sponsored by the National Science Foundation under Grant DMR-080461, Profs. D. Farkas and G. Shiflet grant monitors. The Ni-Al-Mo alloy was processed at NASA Glenn Research Center by Dr. R. D. Noebe. APT measurements were performed at the Northwestern University Center for Atom Probe Tomography (NUCAPT). The LEAP tomograph at NUCAPT was purchased and upgraded with grants from the NSF-MRI (DMR-0420532, Dr. Charles Bouldin, grant officer) and ONR-DURIP (N00014-0400798, N00014-0610539, N00014-0910781, N00014-1712870, Dr. J. Christodoulou, grant officer). NUCAPT received support from the MRSEC program (NSF DMR-1720139) at the Materials Research Center, the SHyNE Resource (NSF ECCS-1542205), and the Initiative for Sustainability and Energy (ISEN) at Northwestern University. We wish to thank Prof. Peter Voorhees for helpful discussions and associate research Prof. Dieter Isheim for managing NUCAPT. We want to thank Ms. Elizaveta Y. Plotnikov for calibration work of the APT, and also thank Dr. Ivan Blum for the edge-to-edge inter-precipitate spacing calculations. NUCAPT received support from the MRSEC program (NSF DMR-1720139) at the Materials Research Center, the SHyNE Resource (NSF ECCS-1542205), and the Initiative for Sustainability and Energy (ISEN) at Northwestern University.

Table 1. Temporal evolution of the nanostructural properties of γ'(L1$_2$)-precipitates determined by atom-probe tomography for a Ni–6.5 Al–9.9 Mo at. % alloy aged at 978 K (705 $^o$C). The threshold value of the isoconcentration surface for the γ(f.c.c.)/γ'(L1$_2$) heterophase-interface, $\eta$ (Al), the number of the identified precipitates, $N_{ppt}$, the γ'(L1$_2$)-precipitate mean radius, $<R(t)>$, number density, $N_v(t)$, volume fraction, $\phi(t)$, the mean edge-to-edge interprecipitate distance, $<\lambda_{edge-edge}>$, and the detected percentage of γ'(L1$_2$)-precipitates interconnected by necks, $f$, are given with their standard errors.

| t(h) | Atoms collected (million) | $\eta$(Al) at. % | $N_{ppt}$ | $<R(t)>$ ±2σ (nm) | $N_v(t)$ ±2σ ×10$^{24}$(m$^{-3}$) | $\phi(t)$ ±2σ (%) | $<\lambda_{edge-edge}>$±2σ (nm) | $f$±2σ (%) |
|---|---|---|---|---|---|---|---|---|
| 0.167 | 81.3 | 11.10 | 1779 | 2.43±0.78 | 1.88±0.04 | 7.76±0.18 | 6.76±3.24 | 21.02±1.02 |
| 0.25 | 56.5 | 11.00 | 1172.5 | 2.51±0.81 | 1.78±0.05 | 8.25±0.24 | 7.01±3.21 | 14.09±1.02 |
| 1 | 37.6 | 10.84 | 316.5 | 3.83±1.22 | 0.755±0.042 | 13.29±0.75 | 10.25±6.25 | 9.1±1.46 |
| 4 | 59.2 | 10.80 | 221 | 5.6±1.74 | 0.292±0.020 | 15.18±1.02 | 12.88±8.38 | 7.4±1.58 |
| 16 | 165.3 | 10.60 | 186 | 8.13±2.56 | 0.0968±0.007 | 16.68±1.22 | 20.45±15.57 | 5.3±1.37 |
| 64 | 437.2 | 10.50 | 136 | 12.08±4.61 | 0.0268±0.002 | 16.87±1.42 | 34.59±25.41 | ND |
| 256 | 1047.3 | 10.45 | 68.5 | 19.12±5.22 | 0.0051±0.003 | 17.68±2.14 | 45.47±39.07 | ND |
| 1024$^a$ | 394.8 | 10.43 | 10.5 | 32.52±15.38 | 0.0011±0.001 | 17.78±5.49 | / | ND |

$^a$There are no bounded precipitates in the LEAP tomography dataset of the specimen aged for 1024 h.

Table 2. The saddle point energy, $E^i_{sp-p,q}$, which is the binding energy of atom $i$ to the saddle point between 1$^{st}$ NN sites $p$ and $q$, the attempt frequency, $\nu^i$, and the vacancy-solute ghost potentials, $\varepsilon_{V-S}$, are listed and described in detail in reference [4].

| | Ni | Al | Mo |
|---|---|---|---|
| $E^i_{sp-p,q}$ (eV) | -9.750 | -9.412 | -10.584 |
| $\nu^i$ (Hz) | $1.10 \times 10^{15}$ | $1.10 \times 10^{15}$ | $5.56 \times 10^{14}$ |
| $\varepsilon_{V-S}$ (eV/atom) | -0.221 | -0.223 | -0.214 |

Table 3. The pair-wise atomic interaction energies (eV) for Ni-Ni, Al-Al, and Mo-Mo from the first-principles calculations utilizing the Chen-Möbius inversion-lattice method [64]. Negative values mean attractive interactions and positive values mean repulsive interactions between atoms.

| $\varepsilon_{i-j}^{k}$ (eV) | Ni-Ni | Al-Al | Mo-Mo | Ni-Al | Ni-Mo | Al-Mo |
|---|---|---|---|---|---|---|
| $1^{st}$ NN | -0.7492 | -0.5788 | -0.7224 | -0.7487 | -0.7492 | -0.6672 |
| $2^{nd}$ NN | -0.0142 | -0.0271 | -0.0205 | 0.0351 | 0.0277 | 0.0245 |
| $3^{rd}$ NN | 0.0135 | 0.0095 | -0.0194 | -0.0288 | 0.00576 | 0.0231 |
| $4^{th}$ NN | -0.0069 | -0.0117 | -0.0096 | 0.0116 | -0.0192 | -0.0615 |

Table 4. Temporal evolution of the concentration of *element i* in the far-field (*ff*) γ(f.c.c.)-matrix ( $C_i^{\gamma,ff}$ ) and the γ'(L1$_2$)-precipitates's concentration ( $C_i^{\gamma'}$ ), and extrapolated ($t = \infty$) concentrations to the solvus curves at 978 K (705 °C). (Unit: at. %)

| 978K | γ(f.c.c.)-matrix(far-field) | | | γ'(L1$_2$)-precipitates(core) | | |
|------|------|------|------|------|------|------|
| t(h) | $C_{Ni}^{\gamma} \pm 2\sigma$ | $C_{Al}^{\gamma} \pm 2\sigma$ | $C_{Mo}^{\gamma} \pm 2\sigma$ | $C_{Ni}^{\gamma'} \pm 2\sigma$ | $C_{Al}^{\gamma'} \pm 2\sigma$ | $C_{Mo}^{\gamma'} \pm 2\sigma$ |
| 0.167 | 84.74±0.03 | 5.01±0.02 | 10.25±0.03 | 74.86±0.18 | 17.81±0.10 | 7.33±0.18 |
| 0.25 | 84.87±0.04 | 4.85±0.02 | 10.28±0.04 | 74.96±0.21 | 17.75±0.12 | 7.29±0.17 |
| 1 | 85.04±0.05 | 4.57±0.03 | 10.39±0.04 | 75.52±0.18 | 17.46±0.12 | 7.03±0.14 |
| 4 | 85.11±0.04 | 4.39±0.02 | 10.49±0.03 | 75.99±0.12 | 17.24±0.09 | 6.77±0.07 |
| 16 | 85.15±0.03 | 4.26±0.02 | 10.59±0.03 | 76.23±0.07 | 17.11±0.09 | 6.66±0.04 |
| 64 | 85.17±0.02 | 4.21±0.01 | 10.62±0.02 | 76.35±0.04 | 17.05±0.07 | 6.60±0.06 |
| 256 | 85.18±0.06 | 4.18±0.03 | 10.64±0.03 | 76.45±0.09 | 16.99±0.09 | 6.56±0.04 |
| 1024 | 85.18±0.06 | 4.16±0.03 | 10.66±0.06 | 76.58±0.09 | 16.91±0.08 | 6.51±0.05 |
| ∞ | 85.21±0.01 | 4.12±0.02 | 10.67±0.01 | 76.64±0.12 | 16.89±0.06 | 6.47±0.06 |
| ThermoCalc | 85.29 | 4.17 | 10.54 | 75.25 | 17.98 | 6.77 |

Table 5. Temporal evolution of the interfacial concentration width, $\delta(t)$, between the γ(f.c.c.)- and γ'(L12)-precipitate-phases, and the relationship between $\delta(t)$ and $\langle R(t) \rangle$ as the alloy is aged. The interfacial compositional width is defined using the 10 and 90% points when fitting the concentration profiles from the proximity histogram for each dataset with a spline curve.

| 978 K t(h) | Width of interface (nm), $\delta_i(t)$ | | | $\delta_i(t)/\langle R(t) \rangle$ | | |
|---|---|---|---|---|---|---|
| | Al | Mo | Ni | Al | Mo | Ni |
| **0.167** | 3.34 | 3.18 | 2.75 | 1.37±0.44 | 1.31±0.42 | 1.13±0.36 |
| **0.25** | 2.85 | 3.05 | 2.62 | 1.14±0.37 | 1.22±0.39 | 1.04±0.34 |
| **1** | 2.46 | 2.85 | 2.46 | 0.64±0.20 | 0.74±0.24 | 0.64±0.20 |
| **4** | 2.35 | 2.77 | 2.38 | 0.42±0.13 | 0.49±0.15 | 0.43±0.13 |
| **16** | 2.19 | 2.71 | 2.31 | 0.27±0.08 | 0.33±0.10 | 0.28±0.09 |
| **64** | 2.14 | 2.68 | 2.28 | 0.18±0.07 | 0.22±0.08 | 0.19±0.07 |
| **256** | 2.12 | 2.65 | 2.11 | 0.11±0.03 | 0.14±0.04 | 0.11±0.03 |
| **1024** | 2.02 | 2.63 | 2.07 | 0.06±0.03 | 0.08±0.04 | 0.06±0.03 |

Figure 1. Horizontal solid-lines mark the average core composition of the $\gamma'(L1_2)$-precipitates (right) and the far-field (*ff*) composition of the $\gamma$(f.c.c.)-matrix (left) measured from a proximity histogram concentration profile obtained from atom-probe tomographic (APT) images of Ni-6.5 Al-9.9 Mo at.% specimens, aged for 64 h at 978 K (705 ºC). Concentration profiles from three separate specimens analyzed employing APT were combined to obtain this concentration profile, which originates from $\gamma$(f.c.c.)/$\gamma'(L1_2)$-heterophase-interfaces of 136 precipitates ($<R(t)> = 12.08 \pm 4.61$ nm, Table 1. The inset figure demonstrates the analytical sensitivity of this method; note the slight retention of Ni atoms adjacent to the $\gamma$(f.c.c.)/$\gamma'(L1_2)$ -heterophase -interface.

Figure 2. A partial ternary phase diagram of the Ni-Al-Mo system at 978 K (705 ºC) calculated using the Grand Canonical Monte Carlo (GCMC) simulation technique, showing the proximity of the Ni-6.5Al-9.9Mo at.% alloy to the ($\gamma$(f.c.c.) plus $\gamma'(L1_2)$/$\gamma$(f.c.c.)) solvus curve. The tie-lines are drawn through the nominal composition of the alloy and between the equilibrium phase-compositions determined by extrapolation of the atom-probe tomographic (APT) concentration data to infinite time. Equilibrium solvus curves determined by Thermo-Calc, using databases for nickel-based superalloys due to Saunders, are superimposed on the GCMC calculated phase diagram for comparative purposes.

Figure 3. Temporal evolution of the $\gamma'(L1_2)$-precipitate-phase nanostructure in Ni-6.5 Al-9.9Mo at.% specimens, aged at 978 K (705 ºC), is revealed within a series of atom-

probe tomographic (APT) images. Each parallelepiped is a $25\times25\times75$ nm$^3$ (46,875 nm$^3$) subset of an analyzed volume and contain approximately 2.1 million atoms per parallelepiped. The $\gamma$(f.c.c.)/$\gamma'$(L1$_2$)-heterophase-interfaces are delineated in red with $\eta$(Al) iso-concentration surfaces.

Figure 4. The temporal evolution of the value of the fraction of $\gamma'$(L1$_2$)-precipitates interconnected by necks, $f(t)$, and the mean inter-precipitate edge-to-edge distance, $<\lambda_{edge-edge}>$.

Figure 5. An atom-probe tomographic (APT) reconstructed volume of a Ni-6.5Al-9.9Mo at.% alloy aged at 978 K (705 °C) for 10 min displaying: (a) the $\gamma'$(L1$_2$)-precipitates delineated by 11.1 at.% Al iso-concentration surfaces; (b)The atoms within $\gamma'$(L1$_2$)-precipitate-pairs, labeled in (a), Ni atoms are shown in green, Mo and Al atoms in purple and red, respectively, demonstrating that the {hkl} lattice planes extend across the necks into $\gamma'$(L1$_2$)-precipitates; and (c) A snapshot of the coagulation and coalescence mechanism extracted from vacancy-mediated lattice-kinetic Monte Carlo (LKMC) simulations at an aging time of 5 min.

Figure 6. The temporal evolution of the $\gamma'$(L1$_2$)-precipitate volume fraction, $\phi(t)$, number density, $N_v(t)$, and mean precipitate radius, $<R(t)>$, for Ni–6.5 Al–9.9Mo at.% aged at 978 K (705 °C). The quantity $<R(t)>$ is proportional to $t^{1/3}$ during quasi-

stationary coarsening for aging times of 16 h and longer, as predicted by the UO, KV and PV coarsening models for ternary alloys. Once the equilibrium volume fraction is approximately achieved after 16 h, the temporal dependence of the quantity $N_v$(t) achieves the $t^{-1}$ prediction of all the coarsening models, UO, KV and PV.

Figure 7. Temporal evolution of the scaled $\gamma'$(L1$_2$)-precipitate size distributions (PSDs) for Ni–6.5 Al–9.9 Mo at.% alloy aged at 978 K (705 °C) for 1/4 to 256 h, with $\phi^{eq} = 18.68\%$ compared to the stationary prediction of the Akiwa-Voorhees (AV) model [24] for $\phi^{eq} = 20\%$. An abundancy of $\gamma'$(L1$_2$)-precipitates are not detected in the 1024 h aging state, since the diameter of the precipitates is large compared to the diameter of a nanotip.

Figure 8. Temporal evolution of the edge-to-edge inter-precipitate distance distribution for a Ni-6.5Al-9.9Mo at.% alloy aged at 978 K (705 °C).

Figure 9. The partitioning ratios, $\kappa_i$, which are defined as the concentration of element $i$ in the $\gamma'$(L1$_2$)-phase, divided by its concentration in the $\gamma$(f.c.c.)-phase [22], where $i =$ Ni, Al or Mo. They demonstrate that Al partitions to the $\gamma'$(L1$_2$)-precipitates, and Ni and Mo to the $\gamma$(f.c.c.)-matrix.

Figure 10. The values of the supersaturations in the $\gamma$(f.c.c.)-matrix, $\left|\Delta C_i^{\gamma}(t)\right|$, and the $\gamma'$(L1$_2$)-precipitates, $\left|\Delta C_i^{\gamma'}(t)\right|$, decrease as $t^{-1/3}$ in the coarsening regime for the Ni-

6.5Al-9.9Mo at.% alloy, as predicted by the UO, KV and PV mean-field models for isothermal stationary coarsening in ternary alloys: see equation (3) in the text. (a) The supersaturation values in the γ(f.c.c.)-matrix, $\left|\Delta C_i^\gamma(t)\right|$, decrease as approximately $t^{1/3}$ in the quasi-coarsening regime; (b) The supersaturation values in the γ'(L1$_2$)-precipitates, $\left|\Delta C_i^{\gamma'}(t)\right|$, decrease as $t^{1/3}$ in the coarsening regime.

Figure 11. Effect of aging time on the interfacial compositional width between the γ(f.c.c.)- and γ'(L1$_2$)-phases. The interfacial compositional width is defined using the 10 and 90% points of the concentration profiles obtained from proximity histograms and fitting each concentration profile from each dataset with a spline curve. The $<R(t)>$ data from Fig. 6 is used to display the relationship between $\delta_i(t)$ and $<R(t)>$ as the alloy is aged. The quantity $\delta_i(t)$ decreases continuously with increasing $<R(t)>$. (a) The effect of aging time on the interfacial compositional width; (b) The effect of aging time on the normalized interfacial compositional width; (c) The relationship between $\delta_i(t)$ and $<R(t)>$ as the alloy is aged. The quantity $\delta_i(t)$ decreases continuously with increasing $<R(t)>$.

Figure 12. Vickers microhardness versus aging time at 978 K (705 ºC) for the Ni-6.5Al-9.9Mo at.% alloy.

Figure 13. Scanning transmission electron microscope (STEM) image of our Ni-6.5Al-9.9Mo at.% alloy aged for 1024 h at 978 K (705 ºC) reveals a uniform

distribution of spheroidal $\gamma'(L1_2)$-precipitates. This image was recorded near the [001]

zone axis.

Figure 1.

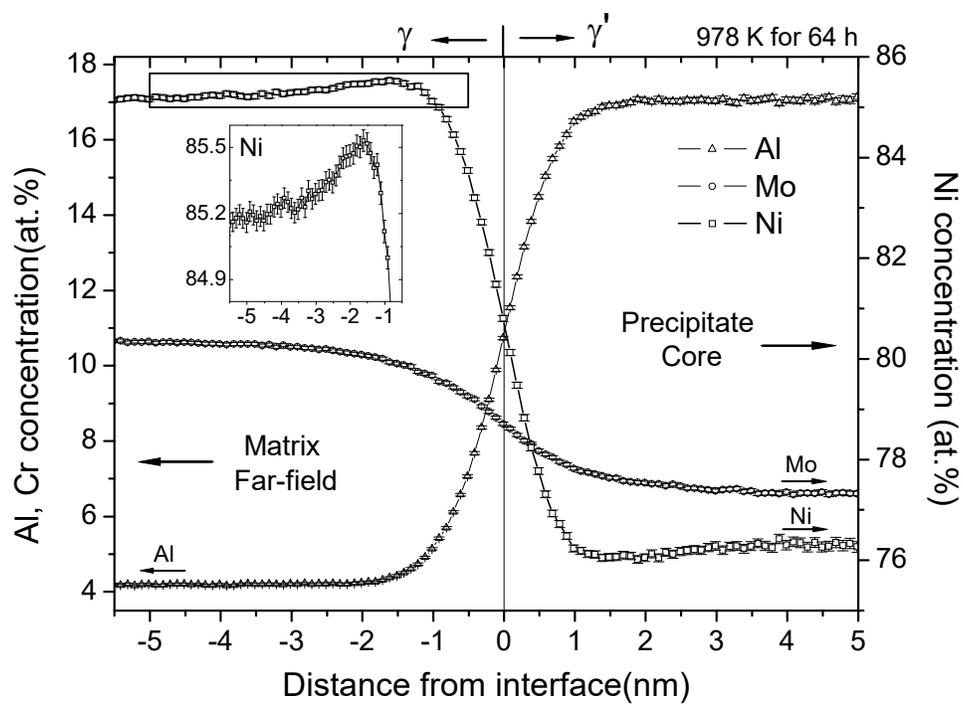

Figure 2.

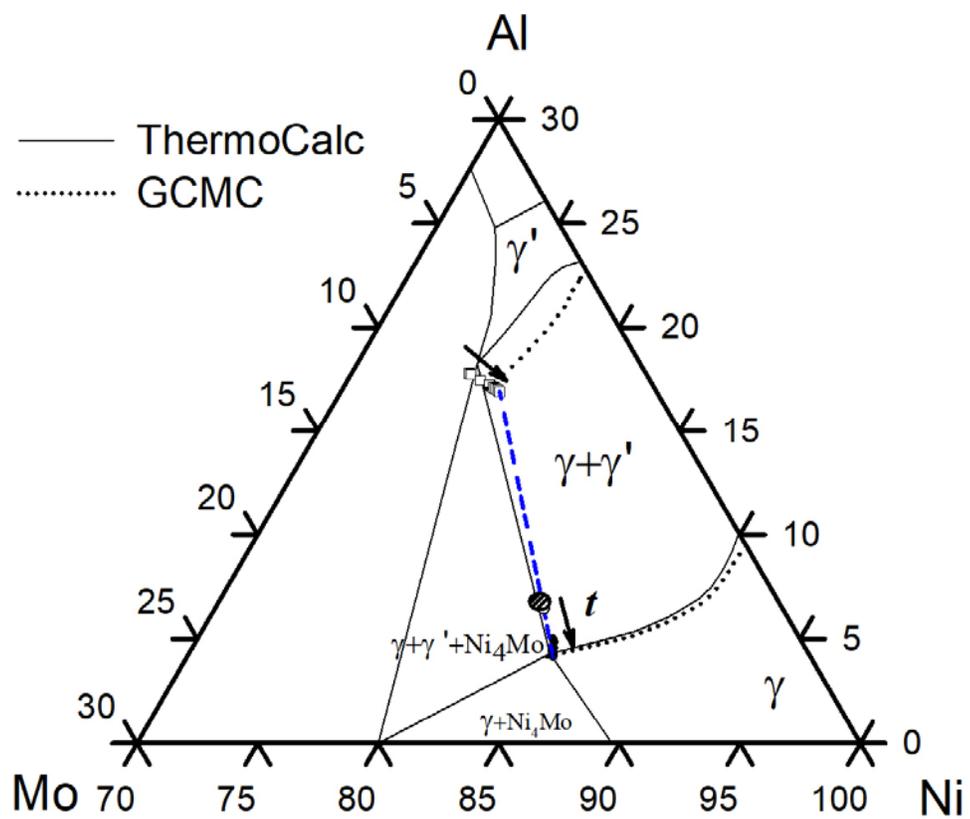

Figure 3.

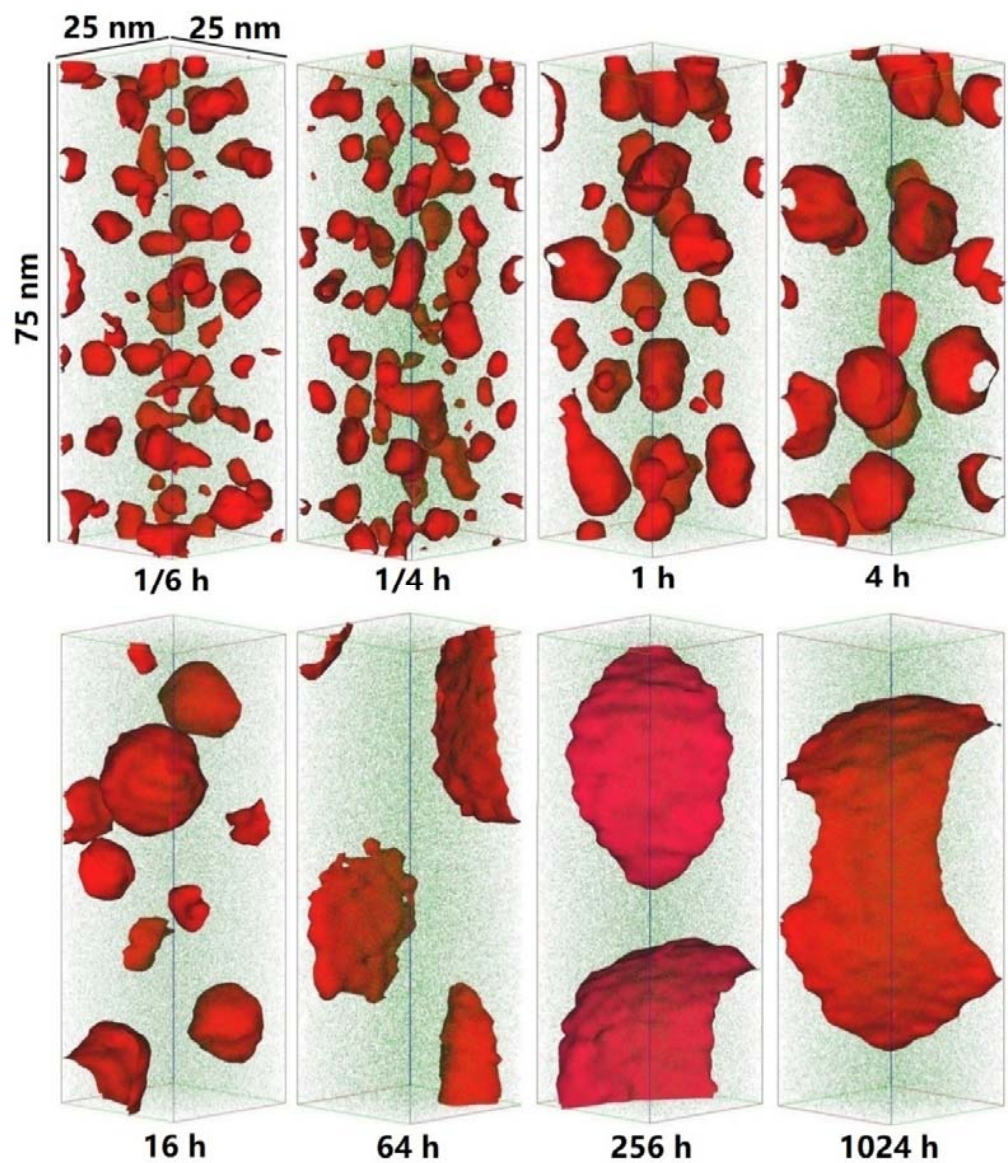

Figure 4.

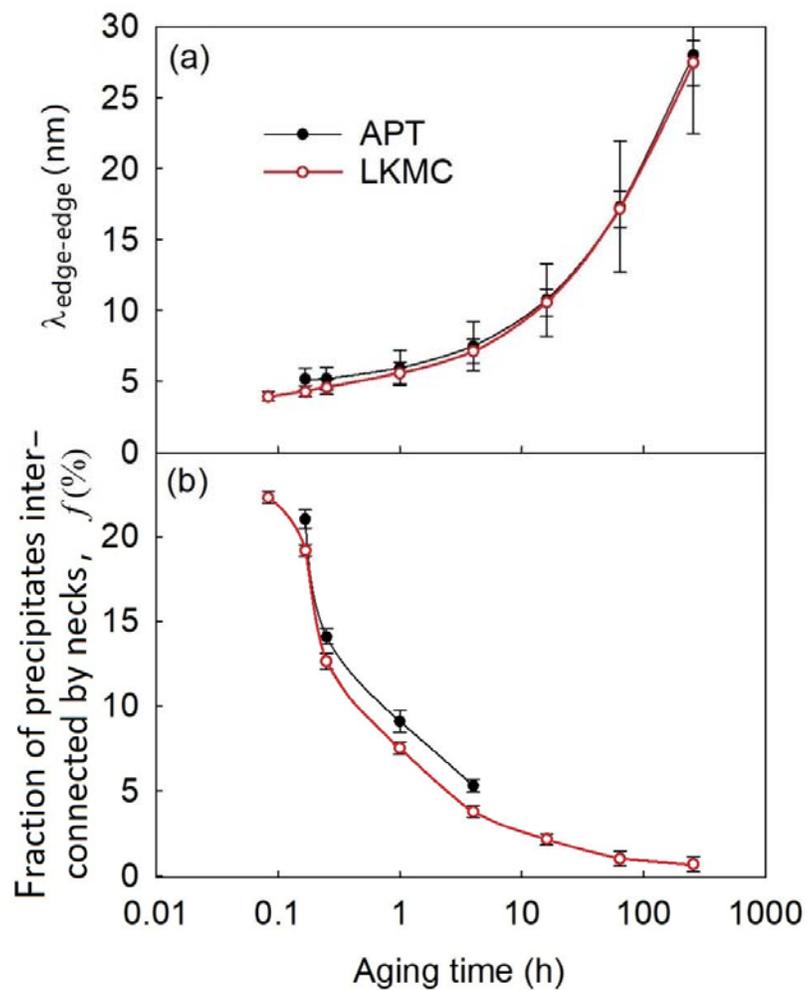

Figure 5.

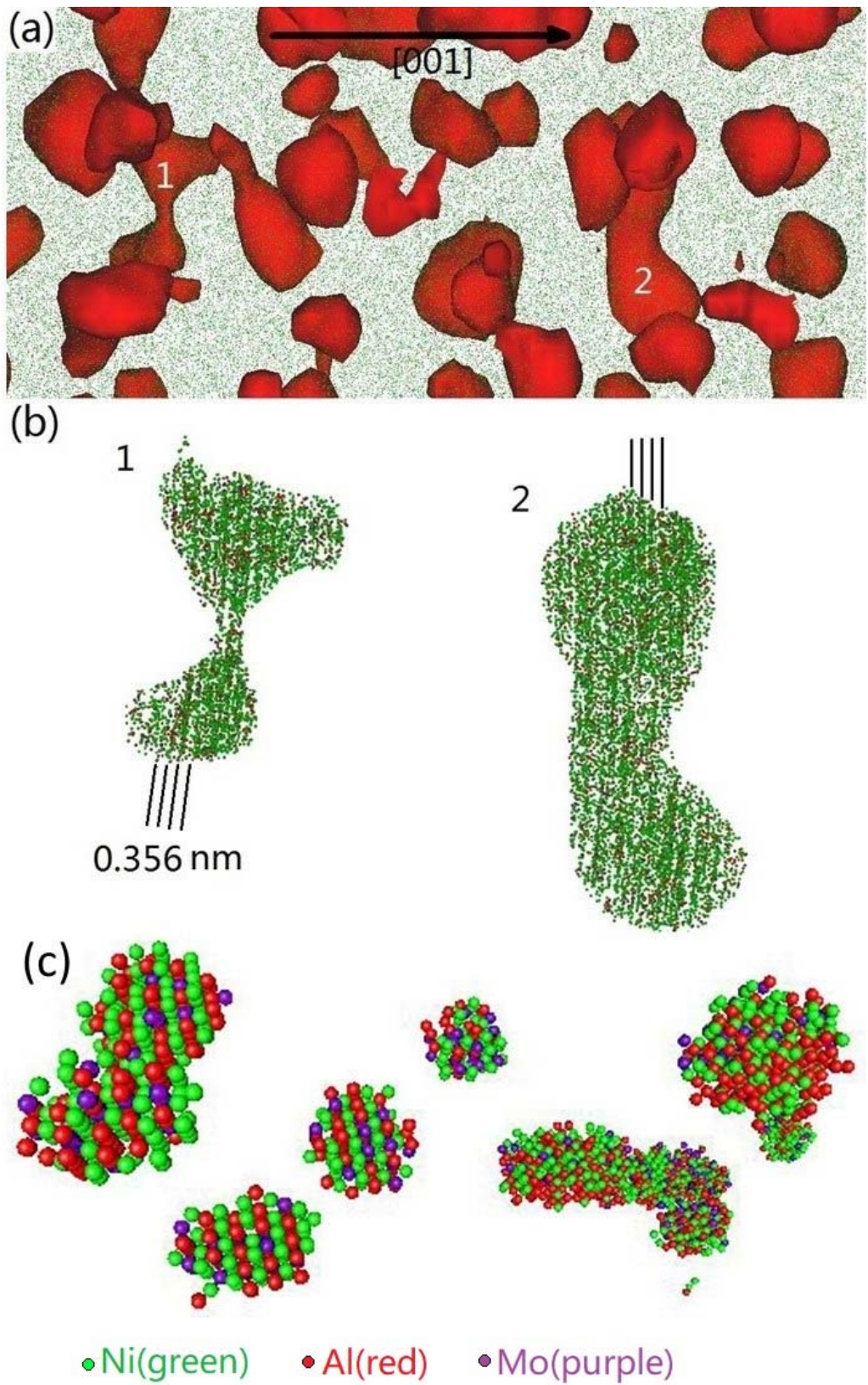

Figure 6.

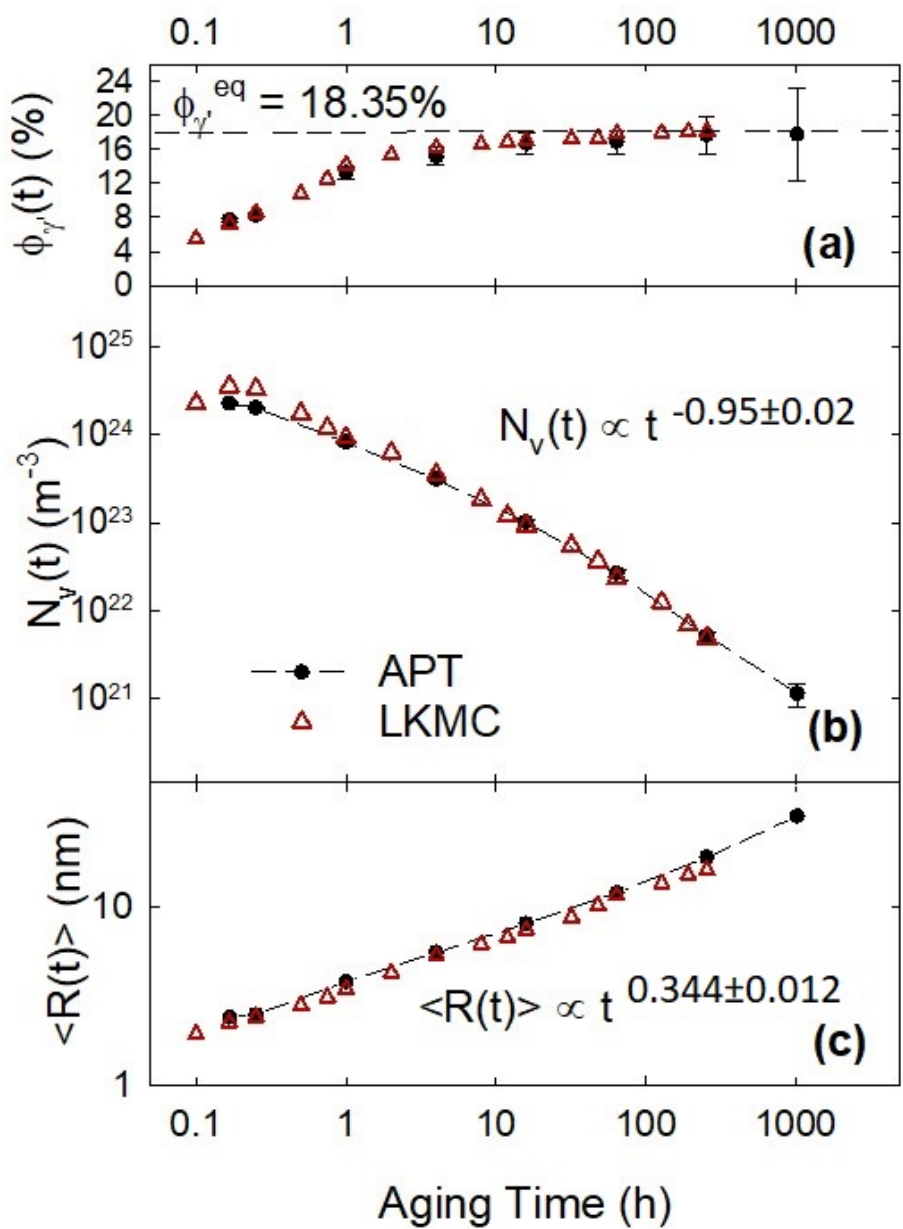

Figure 7.

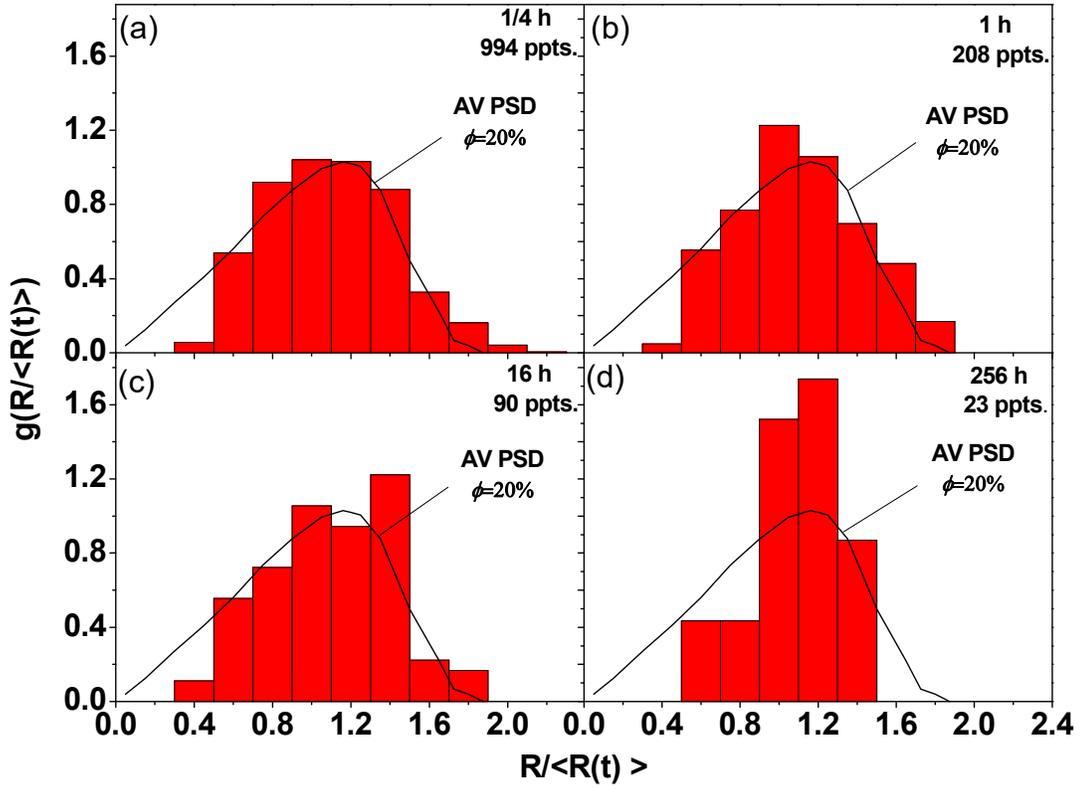

Figure 8.

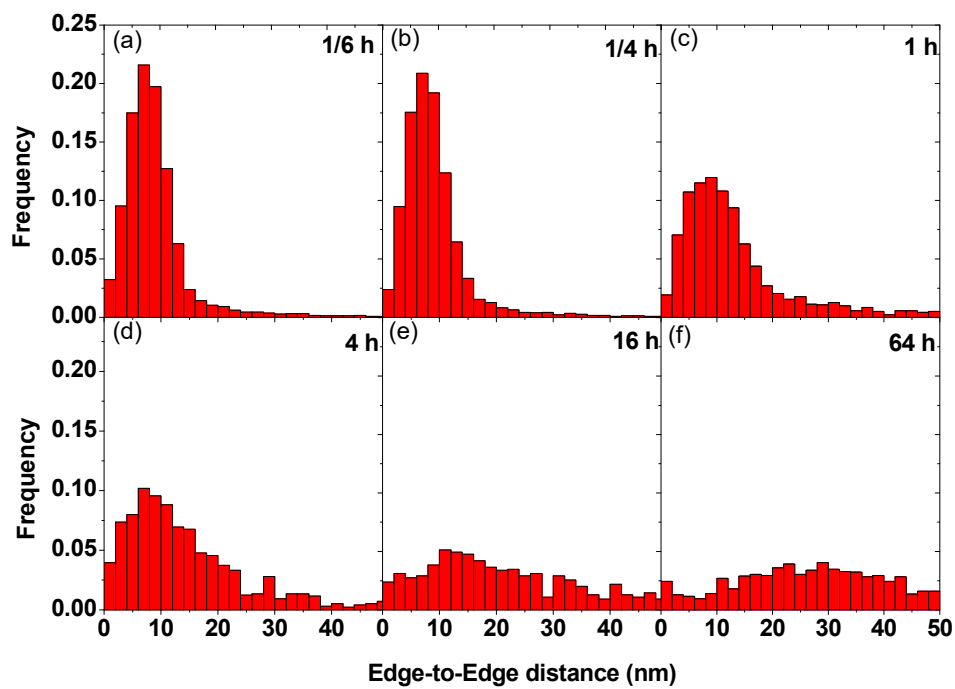

Figure 9.

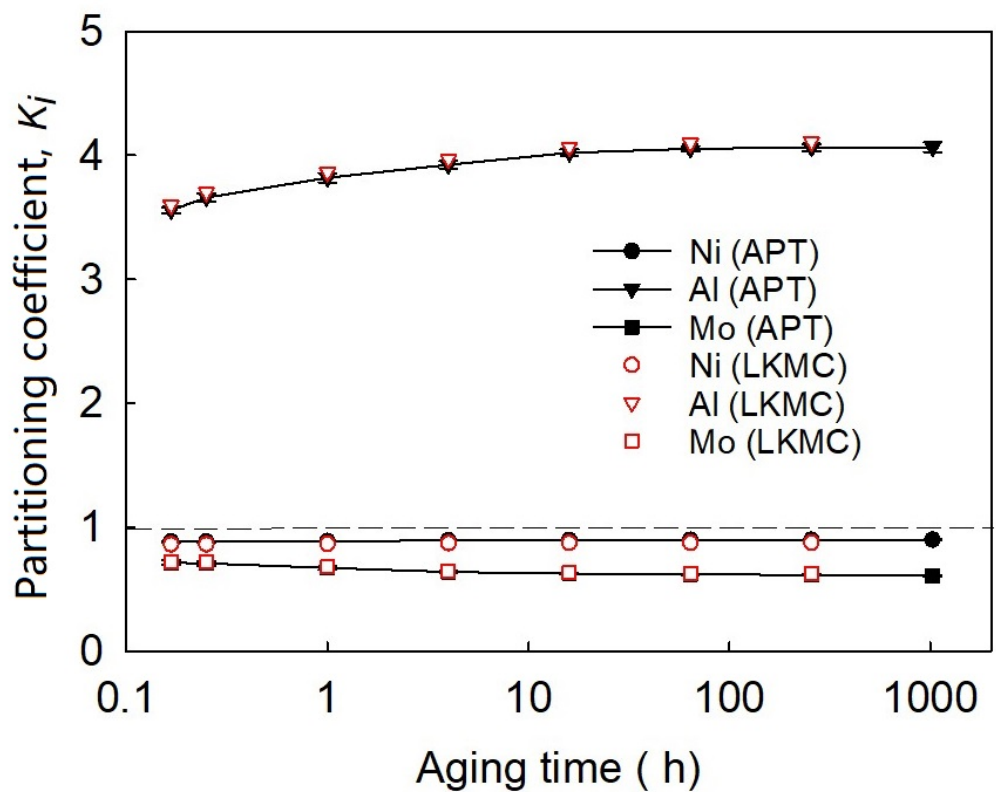

Figure 10(a)

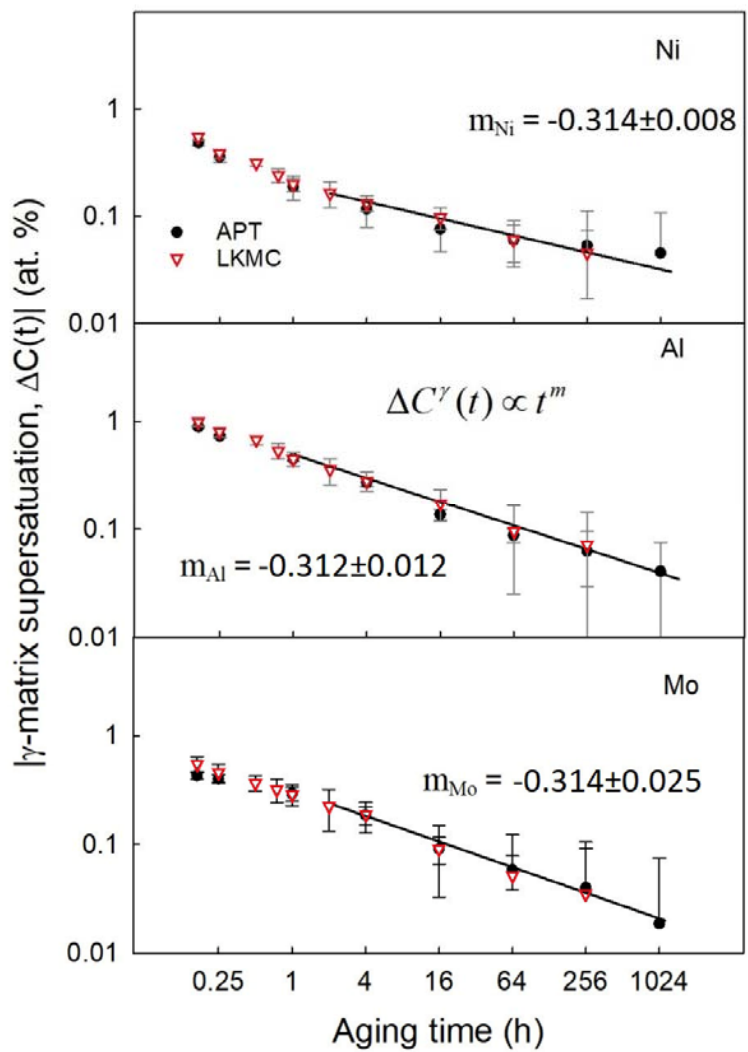

Figure 10(b)

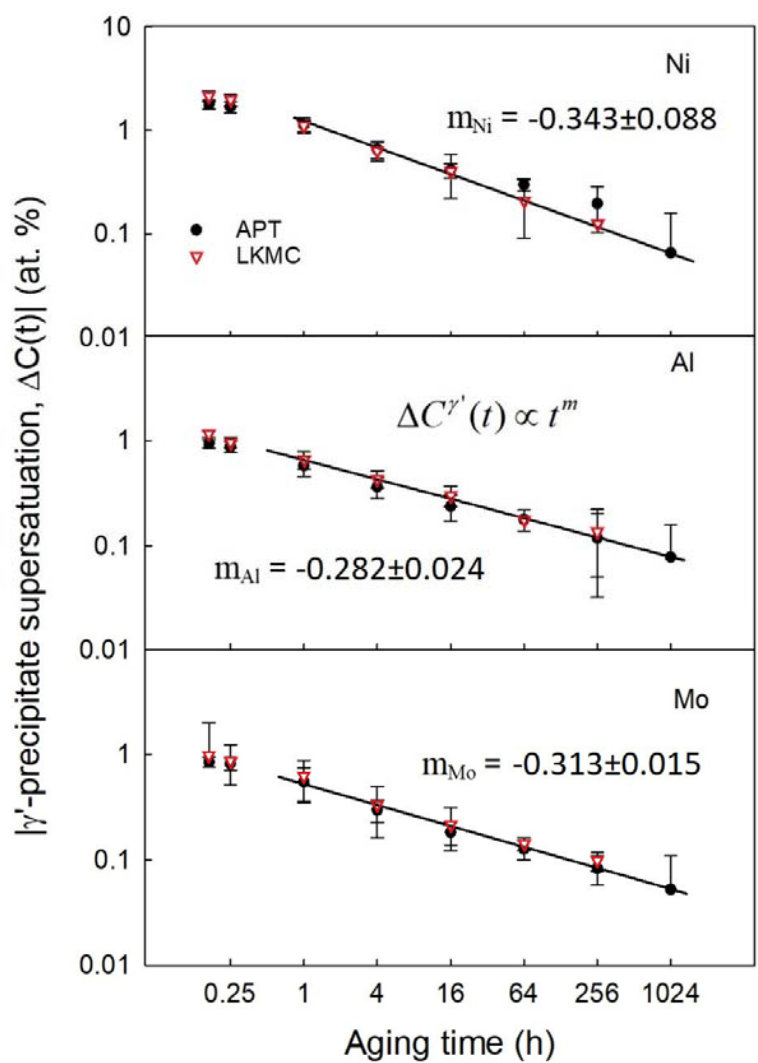

Figure 11(a) & (b)

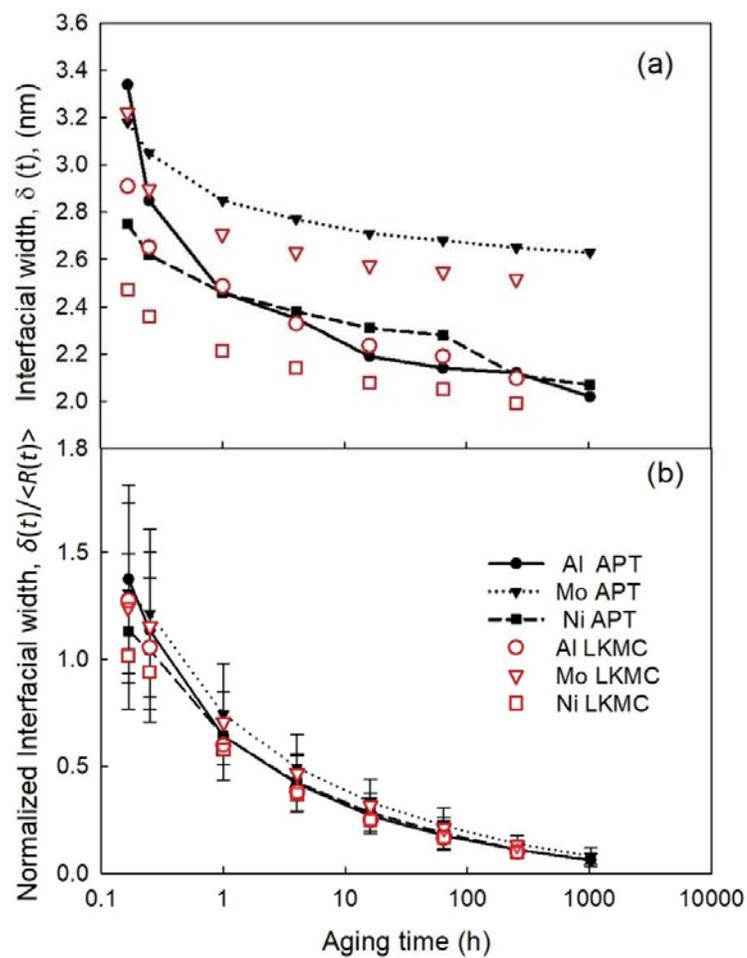

Figure 11(c)

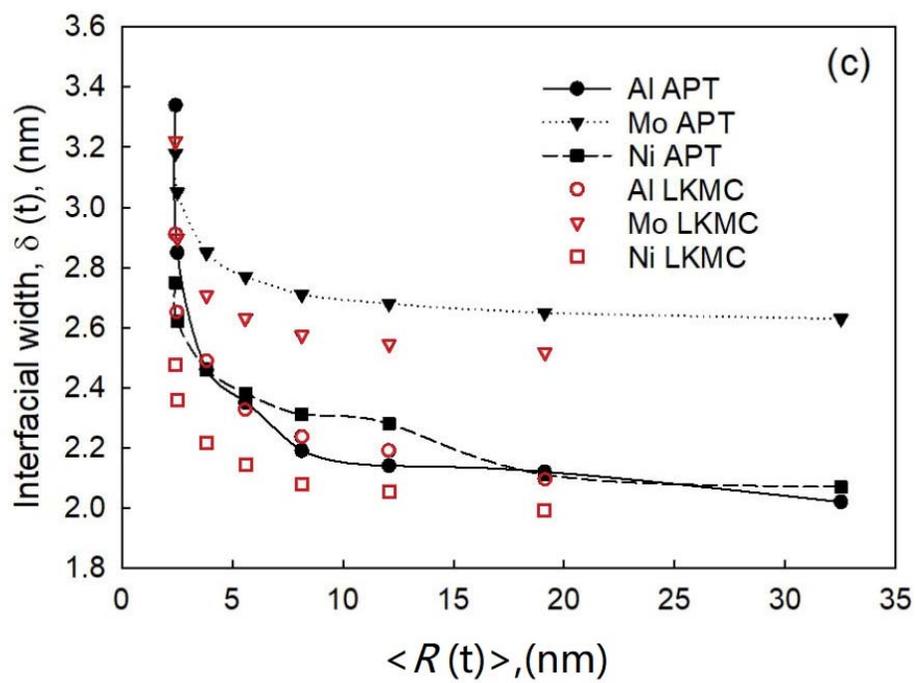

Figure 12.

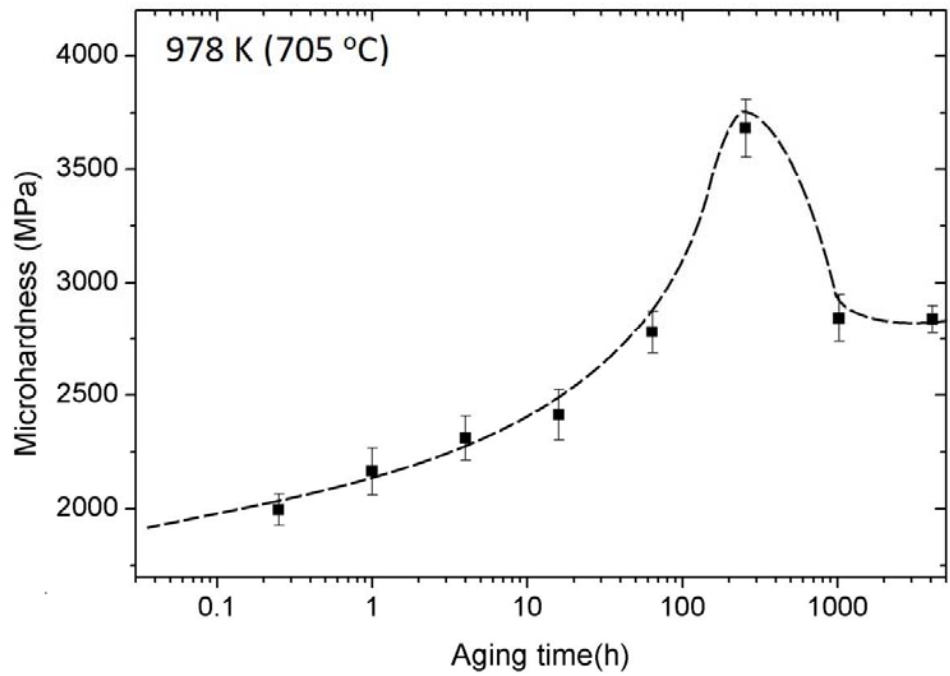

Figure   13.

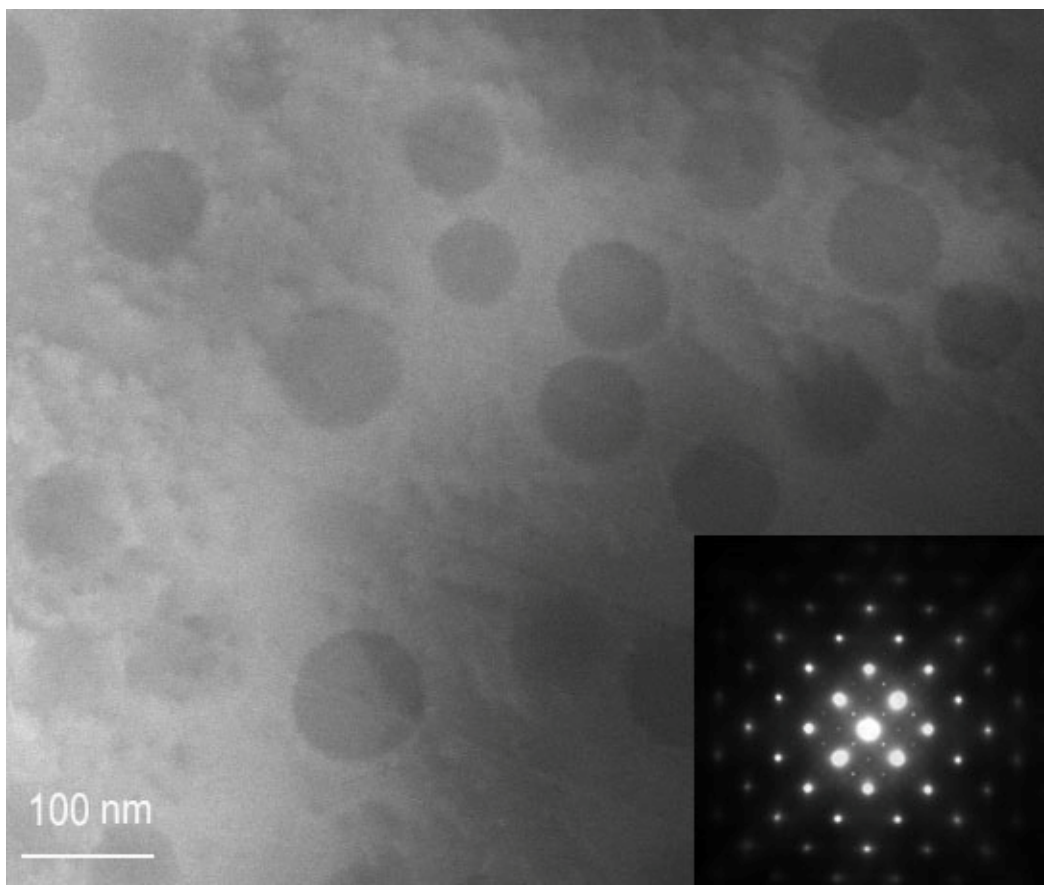